\documentclass[12pt,fleqn,a4paper]{article} 
\usepackage{epsfig,graphics,graphicx}
\usepackage{clshan-math,clshan-dm-dd}
\newcommand{\imageswitch} [2] {#2}
\def \lsim {\:\raisebox{-0.5ex}{$\stackrel{\textstyle<}{\sim}$}\:}
\def \gsim {\:\raisebox{-0.5ex}{$\stackrel{\textstyle>}{\sim}$}\:}
\begin{document}
\thispagestyle{empty}
\begin{flushright}
 March 2009
\end{flushright}
\begin{center}
{\Large\bf
 Determining the Mass of Dark Matter Particles \\ \vspace{0.2cm}
 with Direct Detection Experiments}            \\
\vspace{0.7cm}
 {\sc Chung-Lin Shan} \\
\vspace{0.5cm}
 {\it School of Physics and Astronomy, Seoul Nat'l Univ.,
      Seoul 151-747, Republic of Korea} \\
\vspace{0.1cm}
 {\it E-mail:} {\tt cshan@hep1.snu.ac.kr}
\end{center}
\vspace{1cm}
\begin{abstract}
 In this article I review two data analysis methods
 for determining the mass
 (and eventually the spin--independent cross section on nucleons)
 of Weakly Interacting Massive Particles
 with {\em positive} signals from direct Dark Matter detection experiments:
 a maximum likelihood analysis with {\em only one} experiment
 and a {\em model--independent} method requiring at least two experiments.
 Uncertainties and caveats of these methods
 will also be discussed.
\end{abstract}
\clearpage
\section{Introduction}
 There is strong evidence that
 more than 80\% of all matter in the Universe is dark
 (i.e., interacts at most very weakly
  with electromagnetic radiation and ordinary matter).
 The dominant component of this cosmological Dark Matter
 should be due to some yet to be discovered, non--baryonic particles.
 Weakly Interacting Massive Particles (WIMPs) $\chi$
 arising in several extensions of
 the Standard Model of electroweak interactions
 are one of the leading candidates for Dark Matter.
 WIMPs are stable particles
 with masses roughly between 10 GeV and a few TeV
 and interact with ordinary matter only weakly
 (for reviews
  of WIMPs and some other possible candidates for Dark Matter,
  see Refs.~\cite{{SUSYDM96}, {Bertone05}, {Steffen08}}).

 Currently,
 the most promising method to detect different WIMP candidates
 is the direct detection of the recoil energy
 deposited in a low--background laboratory detector
 by elastic scattering of ambient WIMPs on the target nuclei
 \cite{{DMdet}, {Freese88}, {Lewin96}}%
\footnote{
 Remind that,
 besides many different candidates for WIMPs,
 it is also possible that
 some other particles are
 (theoretically) candidates for Dark Matter.
 For more details about
 these various possible Dark Matter particles
 in many different (exotic) models or scenarios
 as well as the possible methods to detect them,
 see e.g., articles in Parts 1, 2, and 4 of this focus issue.
}.
 The basic expression for the differential event rate
 for elastic WIMP--nucleus scattering is given by \cite{SUSYDM96}:
\beq
   \dRdQ
 = \calA \FQ \int_{\vmin}^{\vmax} \bfrac{f_1(v)}{v} dv
\~.
\label{eqn:dRdQ}
\eeq
 Here $R$ is the direct detection event rate,
 i.e., the number of events
 per unit time and unit mass of detector material,
 $Q$ is the energy deposited in the detector,
 $F(Q)$ is the elastic nuclear form factor,
 $f_1(v)$ is the one--dimensional velocity distribution function
 of the WIMPs impinging on the detector,
 $v$ is the absolute value of the WIMP velocity
 in the laboratory frame.
 The constant coefficient $\calA$ is defined as
\beq
        \calA
 \equiv \frac{\rho_0 \sigma_0}{2 \mchi \mrN^2}
\~,
\label{eqn:calA}
\eeq
 where $\rho_0$ is the WIMP density near the Earth
 and $\sigma_0$ is the total cross section
 ignoring the form factor suppression.
 The reduced mass $\mrN$ is defined by
\beq
        \mrN
 \equiv \frac{\mchi \mN}{\mchi + \mN}
\~,
\label{eqn:mrN}
\eeq
 where $\mchi$ is the WIMP mass and
 $\mN$ that of the target nucleus.
 Finally,
 $\vmin$ is the minimal incoming velocity of incident WIMPs
 that can deposit the energy $Q$ in the detector:
\beq
   \vmin
 = \alpha \sqrt{Q}
\label{eqn:vmin}
\eeq
 with
\beq
        \alpha
 \equiv \sfrac{\mN}{2 \mrN^2}
\~,
\label{eqn:alpha}
\eeq
 and $\vmax$ is related to
 the escape velocity from our Galaxy
 at the position of the Solar system,
 $\vesc$.

 It was found that,
 by using a time--averaged recoil spectrum $dR/dQ$,
 and assuming that no directional information exists,
 the normalized one--dimensional velocity distribution function
 of incident WIMPs, $f_1(v)$, can be solved
 from Eq.(\ref{eqn:dRdQ}) directly as \cite{DMDDf1v}
\beq
   f_1(v)
 = \calN \cbrac{-2 Q \cdot \ddRdQoFQdQ}\Qva
\~,
\label{eqn:f1v_dRdQ}
\eeq 
 where the normalization constant $\calN$ is given by
\beq
   \calN
 = \frac{2}{\alpha} \cbrac{\intz \frac{1}{\sqrt{Q}} \bdRdQoFQ dQ}^{-1}
\~.
\label{eqn:calN_int}
\eeq 
 Note that,
 firstly,
 because $f_1(v)$ in Eq.(\ref{eqn:f1v_dRdQ})
 is the {\em normalized} velocity distribution,
 the normalization constant $\cal N$ here is {\em independent} of
 the constant coefficient $\cal A$ defined in Eq.(\ref{eqn:calA}).
 Secondly,
 the integral in Eq.(\ref{eqn:calN_int})
 goes over the entire physically allowed range of recoil energies:
 starting at $Q = 0$,
 and the upper limit of the integral has been written as $\infty$.
 However,
 it is usually assumed that
 the WIMP flux on the Earth is negligible
 at velocities exceeding the escape velocity $\vesc$.
 This leads thus to a kinematic maximum of the recoil energy
\beq
   Q_{\rm max, kin}
 = \frac {\vesc^2} {\alpha^2}
\~.
\label{eqn:Qmaxkin}
\eeq

 The velocity distribution function of halo WIMPs
 reconstructed by Eq.(\ref{eqn:f1v_dRdQ})
 is independent of the local WIMP density $\rho_0$
 as well as of the WIMP--nucleus cross section $\sigma_0$.
 However,
 not only the overall normalization constant $\calN$
 given in Eq.(\ref{eqn:calN_int}),
 but also the shape of the velocity distribution,
 through the transformation $Q = v^2 / \alpha^2$ in Eq.(\ref{eqn:f1v_dRdQ}),
 depends on the WIMP mass $\mchi$
 involved in the coefficient $\alpha$ defined in Eq.(\ref{eqn:alpha}).
 In fact,
 any (assumed) value of $\mchi$
 will lead to a well--defined, normalized distribution function $f_1(v)$
 when one uses Eq.(\ref{eqn:f1v_dRdQ}).
 Hence,
 $\mchi$ can be extracted from a {\em single} recoil spectrum
 {\em only if} one makes some assumptions
 about the velocity distribution $f_1(v)$.
 In contrast,
 by comparing two (or more) velocity distributions
 reconstructed from different recoil spectra
 with different target nuclei,
 one could avoid using these assumptions
 and estimate the WIMP mass model--independently.

 The remainder of this article is organized as follows.
 In Sec.~2
 I first review a method for determining the WIMP mass
 with only one direct detection experiment.
 In Sec.~3
 I present a model--independent method for determining $\mchi$
 by combining two experimental data sets.
 Numerical results based on Monte Carlo simulations of future experiments
 and uncertainties and caveats of these two methods
 will also be discussed.
 I conclude in Sec.~4.
 Some technical details for the data analysis
 will be given in an appendix.
\section{With one experiment}
 In this section
 I review the method for determining the WIMP mass
 with only one direct detection experiment
 based on a maximum likelihood analysis
 \cite{{Jackson06}, {Schnee06a}, {Green-mchi07}, {Green-mchi08}}.
\subsection{Maximum likelihood analysis}
 I first describe briefly some (standard) theoretical models/assumptions
 for fitting the elastic WIMP--nucleus scattering spectrum
 to experimental data.
 Then I discuss the determination of the WIMP mass 
 by a maximum likelihood analysis.
 Note here that
 only the most commonly used models/assumptions
 are described as examples
 to show which information is required
 for the maximum likelihood analysis;
 however,
 it should be understood that
 other models or assumptions can also be used.
\subsubsection{Simple model distributions}
 The simplest semi--realistic model halo is a Maxwellian halo.
 The one--dimensional velocity distribution function
 in the rest frame of our Galaxy
 can be expressed as \cite{{Lewin96}, {SUSYDM96}, {DMDDf1v}}
\beq
   f_{1, \Gau}(v)
 = \cleft{\renewcommand{\arraystretch}{1.4}
          \begin{array}{l l l}
          N_\Gau
          v^2\!
          \abrac{  e^{-v^2     / v_0^2}
                 - e^{-\vesc^2 / v_0^2}}\~, & ~~~~~~ &
          {\rm for}~v \le \vesc\~, \\
          0\~, & ~~~~~~ &
          {\rm for}~v > \vesc\~. \\
          \end{array}}
\label{eqn:f1v_Gau_ast}
\eeq
 Here $v_0 \simeq 220$ km/s
 is the orbital velocity of the Sun in the Galactic frame,
 and
\beq
   N_\Gau
 = \bbrac{  \afrac{\sqrt{\pi} v_0^3}{4} \erf\afrac{\vesc}{v_0}
          - \vesc \abrac{\frac{v_0^2}{2} + \frac{\vesc^2}{3}} \~ e^{-\vesc^2 / v_0^2}}^{-1}
\label{eqn:N_Gau_ast}
\eeq
 is the normalization constant which satisfies
\beq
   \int_0^{\vesc} f_1(v) \~ dv
 = 1
\~.
\label{eqn:nor_f1v_vesc}
\eeq
 Note that
 the second term on the right--hand side of Eq.(\ref{eqn:f1v_Gau_ast})
 has been introduced to keep
 the velocity distribution continuous at $v = \vesc$.
 Substituting Eq.(\ref{eqn:f1v_Gau_ast}) into Eq.(\ref{eqn:dRdQ}),
 the integral over the velocity distribution function
 can be calculated as
\beq
   \int_{\vmin}^{\vesc} \bfrac{f_{1, \Gau}(v)}{v} dv
 = N_\Gau
   \afrac{v_0^2}{2}
   \bbrac{  e^{-\alpha^2 Q / v_0^2}
          - \afrac{v_0^2 + \vesc^2 - \alpha^2 Q}{v_0^2} e^{-\vesc^2 / v_0^2}}
\~,
\label{eqn:int_f1v_Gau_ast}
\eeq
 where $\vmin = \alpha \sqrt{Q}$ in Eq.(\ref{eqn:vmin})
 has been used.
 Note that,
 in the $\vesc \to \infty$ limit,
 $N_\Gau \to 4 / \sqrt{\pi} v_0^3$ and
 the integral approaches to
 $(2 / \sqrt{\pi} v_0) \~ e^{-\alpha^2 Q / v_0^2}$.

 On the other hand,
 when we take into account
 the orbital motion of the Solar system around the Galaxy
 as well as that of the Earth around the Sun,
 the velocity distribution function should be modified to
 \cite{{Lewin96}, {SUSYDM96}, {DMDDf1v}}
\beq
   f_{1, \sh}(v)
 = N_\sh v
   \cBigg{  \bbigg{  e^{-(v     - v_{\rm e})^2 / v_0^2}
                   - e^{-(v     + v_{\rm e})^2 / v_0^2}}
          - \bbigg{  e^{-(\vesc - v_{\rm e})^2 / v_0^2}
                   - e^{-(\vesc + v_{\rm e})^2 / v_0^2}} }
\~,
\label{eqn:f1v_sh_ast}
\eeq
 for $v \le \vesc$,
 with the normalization constant
\beqn
        N_\sh
 \=     \cleft{    \frac{\sqrt{\pi} v_{\rm e} v_0}{2}
                   \bbigg{  \erf{\T \afrac{\vesc + v_{\rm e}}{v_0}}
                          + \erf{\T \afrac{\vesc - v_{\rm e}}{v_0}} } }
        \non\\
 \conti ~~~~~~~~~~~~~~~~~~~~ 
        \cright{+ \afrac{v_0^2 + \vesc^2}{2} 
                  \bbigg{  e^{-(\vesc + v_{\rm e})^2 / v_0^2}
                         - e^{-(\vesc - v_{\rm e})^2 / v_0^2} } }^{-1}
\~.
\label{eqn:N_sh_ast}
\eeqn
 Here $v_{\rm e}$ is the Earth's velocity in the Galactic frame
 \cite{{Freese88}, {SUSYDM96}, {Bertone05}}:
\beq
   v_{\rm e}(t)
 = v_0 \bbrac{1.05 + 0.07 \cos\afrac{2 \pi (t - t_{\rm p})}{1~{\rm yr}}}
\~;
\label{eqn:ve}
\eeq
 $t_{\rm p} \simeq$ June 2nd is the date
 on which the velocity of the Earth
 relative to the WIMP halo is maximal.
 Consequently,
 an analytic form of the integral over this velocity distribution
 can be given as
\beqn
 \conti \int_{\vmin}^{\vesc} \bfrac{f_{1, \sh}(v)}{v} dv
        \non\\
 \=     N_\sh
        \cleft{  \frac{\sqrt{\pi} v_0}{2}
                 \cBiggl{ \bbigg{ \erf{\T\afrac{\alpha \sqrt{Q} + v_{\rm e}}{v_0}}
                                 -\erf{\T\afrac{\alpha \sqrt{Q} - v_{\rm e}}{v_0}} } }
                 \cBiggr{-\bbigg{ \erf{\T\afrac{\vesc           + v_{\rm e}}{v_0}}
                                 -\erf{\T\afrac{\vesc           - v_{\rm e}}{v_0}} } } }
        \non\\
 \conti ~~~~~~~~~~~~~~~~ 
        \cBiggr{+ \aBig{\vesc - \alpha \sqrt{Q} \~ }
                  \bbigg{  e^{-(\vesc + v_{\rm e})^2 / v_0^2}
                         - e^{-(\vesc - v_{\rm e})^2 / v_0^2} }}
\~.
\label{eqn:int_f1v_sh_ast}
\eeqn
 For practical, numerical uses,
 an approximate form of the integral over $f_1(v)$
 was introduced as \cite{Lewin96}
\beq
   \int_{\vmin}^{\vesc} \bfrac{f_1(v)}{v} dv
 = c_0 \afrac{2}{\sqrt{\pi} v_0} e^{-\alpha^2 Q/c_1 v_0^2}
\~,
\label{eqn:int_f1v_cc}
\eeq
 where $c_0$ and $c_1$ are two fitting parameters of order unity.
 Not surprisingly,
 their values depend on
 the Galactic orbital and escape velocities,
 the target nucleus,
 the threshold energy of the experiment,
 as well as on the mass of incident WIMPs.
 Note that,
 the characteristic energy
\beq
        Q_{\rm ch}
 \equiv \frac{c_1 v_0^2}{\alpha^2}
\label{eqn:Q_ch}
\eeq
 and thus the shape of the recoil spectrum
 depend highly on the WIMP mass:
 for light WIMPs ($\mchi \ll m_{\rm N}$),
 $Q_{\rm ch} \propto \mchi^2$ and
 the recoil spectrum drops sharply with increasing recoil energy,
 while for heavy WIMPs ($\mchi \gg m_{\rm N}$),
 $Q_{\rm ch} \sim$ const.~and the spectrum becomes flatter.
\subsubsection{Local WIMP density}
 Currently,
 the most commonly used value for
 the local WIMPs density in Eq.(\ref{eqn:calA})
 is given as \cite{{SUSYDM96}, {Bertone05}}
\beq
         \rho_0
 \approx 0.3~{\rm GeV / cm^3}
\~.
\label{eqn:rho0}
\eeq
 However,
 so far it can be estimated only by means of
 the measurement of the rotational velocity of our Galaxy.
 Due to our location inside the Milky Way,
 it is more difficult to measure the accurate rotation curve of our own Galaxy
 than those of other galaxies.
 Thus an uncertainty of around a factor of 2
 has been usually adopted \cite{{SUSYDM96}, {Bertone05}}:%
\beq
   \rho_0
 = 0.2 - 0.8~{\rm GeV / cm^3}
\~.
\label{eqn:rho0_range}
\eeq
\subsubsection{Spin--independent WIMP--nucleus cross section}
 In most theoretical models,
 the spin--independent (SI) WIMP interaction on a nucleus
 with an atomic mass number $A \gsim 30$ dominates
 the spin--dependent (SD) interaction \cite{{SUSYDM96}, {Bertone05}}.
 Additionally,
 for the lightest supersymmetric neutralino,
 which is perhaps the best motivated WIMP candidate
 \cite{{SUSYDM96}, {Bertone05}},
 and for all WIMPs which interact primarily through Higgs exchange,
 the SI scalar coupling is approximately the same
 on both protons p and neutrons n.
 The ``pointlike'' cross section $\sigma_0$ in Eq.(\ref{eqn:calA})
 can thus be written as
\beq
   \sigma_0
 = A^2 \afrac{m_{\rm r, N}}{m_{\rm r, p}}^2 \sigmapSI
\~,
\label{eqn:sigma0SI}
\eeq
 where
\beq
   \sigmapSI
 = \afrac{4}{\pi} m_{\rm r, p}^2 |f_{\rm p}|^2
\~,
\label{eqn:sigmapSI}
\eeq
 and $f_{\rm p}$ is the effective $\chi \chi {\rm p p}$ four--point coupling,
 $A$ is the atomic mass number of the target nucleus.
\subsubsection{Nuclear form factor}
 For the SI cross section,
 an analytic nuclear form factor can be used.
 The simplest one is the exponential form factor,
 first introduced by Ahlen {\it et al.} \cite{Ahlen87}
 and Freese {\it et al.} \cite{Freese88}:
\beq
   F_{\rm ex}^2(Q)
 = e^{-Q / Q_0}
\~.
\label{eqn:FQ_SI_ex}
\eeq
 Here $Q$ is the recoil energy
 transferred from the incident WIMP to the target nucleus,
\beq
   Q_0
 = \frac{1.5}{\mN R_0^2}
\label{eqn:Q0}
\eeq
 is the nuclear coherence energy and
\beq
   R_0
 = \bbrac{0.3 + 0.91 \afrac{\mN}{\rm GeV}^{1/3}}~{\rm fm}
\label{eqn:R0}
\eeq
 is the radius of the nucleus.
 The exponential form factor implies
 a Gaussian form of the radial density profile of the nucleus.
 This Gaussian density profile is simple,
 but not very realistic.
 Engel has therefore suggested
 a more accurate form factor \cite{Engel91},
 inspired by the Woods-Saxon nuclear density profile
 \cite{{SUSYDM96}, {Bertone05}},
\beq
   F_{\rm WS}^2(Q)
 = \bfrac{3 j_1(q R_1)}{q R_1}^2 e^{-(q s)^2}
\~.
\label{eqn:FQ_SI_WS}
\eeq
 Here $j_1(x)$ is a spherical Bessel function,
\beq
   q
 = \sqrt{2 m_{\rm N} Q}
\label{eqn:qq}
\eeq
 is the transferred 3-momentum,
\beq
   R_1
 = \sqrt{R_A^2 - 5 s^2}
\label{eqn:R1}
\eeq
 is the effective nuclear radius%
\footnote{
 In the literature,
 the form factor given in Eq.(\ref{eqn:FQ_SI_WS})
 is also known as the ``Helm'' form factor with
 \cite{{Helm56}, {Lewin96}}
\beq
   R_1
 = \sqrt{R_A^2 + {\T \afrac{7}{3}} \pi^2 r_0^2 - 5 s^2}
\~,
\label{eqn:R1_Helm}
\eeq
 where
\beq
        R_A
 \simeq \abig{1.23 \~ A^{1/3} - 0.6}~{\rm fm},
        ~~~~~~~~~~~~~~ 
%
        r_0
 \simeq 0.52~{\rm fm},
        ~~~~~~~~~~~~~~ 
%
        s
 \simeq 0.9~{\rm fm}.
\label{eqn:RA_r0_ss_Helm}
\eeq
}
 with
\footnote{
 For $R_1$ given by Eq.(\ref{eqn:R1})
 with $s \simeq 1$ fm,
 a more precise approximation for $R_A$
 has also been given \cite{{Eder68}, {Lewin96}}:
\beq
        R_A
 \simeq \abig{1.15 \~ A^{1/3} + 0.39}~{\rm fm}.
\label{eqn:RA_Eder}
\eeq
}
\beq
        R_A
 \simeq 1.2 \~ A^{1/3}~{\rm fm},
\label{eqn:RA}
\eeq
 and
\beq
        s
 \simeq 1~{\rm fm}
\label{eqn:ss}
\eeq
 is the nuclear skin thickness.
\subsubsection{Extended likelihood function}
 Now we are ready to put all pieces
 for predicting the elastic WIMP--nucleus scattering spectrum together
 and then fit this spectrum to experimental data
 by maximizing the logarithm of the extended likelihood function
 \cite{Green-mchi07}:
\beq
   {\cal L}
 = \frac{\lambda^{N_{\rm tot}} \~ e^{-\lambda}}{N_{\rm tot}!} \cdot
   \frac{1}{R} \prod_{a = 1}^{N_{\rm tot}} \adRdQ_{Q = Q_a}
\~.
\label{eqn:calL}
\eeq
 Here
\beq
   \lambda
 = \calE \int_{\Qmin}^{\Qmax} \adRdQ dQ
\label{eqn:lambda_L}
\eeq
 is the expected event number
 with the (assumed) exposure of the experiment, $\calE$,
 $N_{\rm tot}$ is the total number of events
 recorded in one (simulated) experiment,
 $Q_a$ are measured recoil energies in the data set
 between the minimal and maximal cut--off energies,
 $\Qmin$ and $\Qmax$,
 and
\beq
   R
 = \int_{\Qmin}^{\Qmax} \adRdQ dQ
\label{eqn:R}
\eeq
 is the total event rate.

 Note that,
 firstly,
 the definition of $\cal L$ in Eq.(\ref{eqn:calL})
 takes into account the fact that
 the event number $N_{\rm tot}$ and the measured recoil spectrum $\calE (dR / dQ)$
 of each (simulated) experiment are not fixed.
 Secondly,
 except $c_0$ and $c_1$ in Eq.(\ref{eqn:int_f1v_cc}),
 there are {\em two} fitting parameters
 in the extended likelihood function $\cal L$,
 i.e., the WIMP mass $\mchi$ (involved in $\alpha$)
 and the SI WIMP--proton cross section $\sigmapSI$.
\subsection{Numerical results}
 Here I show some numerical results
 with 10,000 simulated experiments
 based on Monte Carlo simulations
 performed by A.~Green \cite{{Green-mchi07}, {Green-mchi08}}.
 $\rmXA{Ge}{76}$ has been chosen as the target nucleus
 with a threshold energy of 10 keV.
 A three--dimensional Maxwellian velocity distribution
 in the Galactic rest frame for an isotropic isothermal WIMP halo,
 taking into account the Earth's motion around the Sun
 with $v_0 = 220$ km/s and $\vesc = 540$ km/s,
 and the Helm form factor
 in Eqs.(\ref{eqn:FQ_SI_WS}), (\ref{eqn:qq}),
 (\ref{eqn:R1_Helm}), and (\ref{eqn:RA_r0_ss_Helm})
 have been used.
 The standard assumption for the local WIMP density of 0.3 GeV/$\rm cm^3$
 has been adopted.

 Note that
 the simulations demonstrated here
 as well as in the next section
 for the method combining two experimental data sets
 are based on several simplified assumptions%
\footnote{
 More realistic modelling with
 e.g.,~other WIMP velocity distributions
 and/or different nuclear form factors
 could in principle be incorporated
 into the maximum likelihood analysis.
}.
 Firstly,
 the sample to be analyzed contains only signal events,
 i.e., is free of background.
 Active background suppression techniques
 \cite{{Baudis07a}, {Drees08}, {Bednyakov08}}%
\footnote{
\label{NJP-DM-Part3}
 For more experimental details about
 current direct detection techniques and
 the next generation detectors,
 see articles in Part 3 of this focus issue.
}
 should make this condition possible.
 Secondly,
 all experimental systematic uncertainties
 as well as the uncertainty on the measurement of the recoil energy
 have been ignored.
 The energy resolution of most existing detectors is so good
 that its error can be neglected 
 compared to the statistical uncertainty for the foreseeable future.
\subsubsection{Statistical uncertainty}
\begin{figure}[t!]
\begin{center}
\imageswitch{
\begin{picture}(16,6.4)
\put(0,0){\framebox(16,6.4){Green-sigmapSI-mchi}}
\end{picture}}
{
\includegraphics[width=16cm]{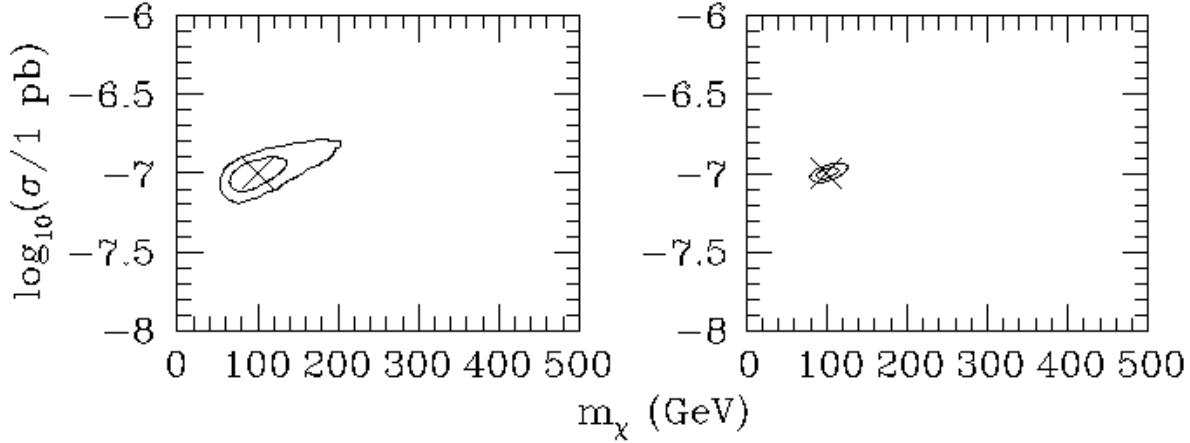}
\vspace*{-0.25cm}
}
\end{center}
\caption{
 Distributions of the best--fit WIMP mass
 and SI WIMP--proton cross section
 on the cross section versus WIMP mass plane.
 The input WIMP mass and the cross section are
 100 GeV and $10^{-7}$ pb,
 respectively.
 The exposures have been assumed to be
 $3 \times 10^3$ (left) and $3 \times 10^4$ (right) kg-day
 and the corresponding expected event numbers are 78 and 780,
 respectively.
 In each frame,
 the contours contain 68\% and 95\% of the simulated experiments.
 See the text for further details
 (Plots from \cite{Green-mchi07}).
}
\label{fig:Green_sigmapSI_mchi}
\end{figure}
 Figs.~\ref{fig:Green_sigmapSI_mchi} show
 the distributions of the best--fit WIMP mass $\mchi$
 and SI WIMP--proton cross section $\sigmapSI$
 on the cross section versus WIMP mass plane.
 The input WIMP mass and the cross section are 100 GeV and $10^{-7}$ pb,
 respectively.
 The exposures have been assumed to be
 $3 \times 10^3$ (left) and $3 \times 10^4$ (right) kg-day
 and the corresponding expected event numbers are 78 and 780%
\footnote{
 Since the event number is directly proportional to
 the product of the cross section $\sigmapSI$ and the exposure $\calE$,
 it is equivalent to assume $\sigmapSI = 10^{-8}$ pb
 and exposures of $3 \times 10^4$ and $3 \times 10^5$ kg-day.
},
 respectively.
 It can be seen that,
 especially for the smaller exposure,
 the distribution is {\em asymmetric} and
 there are (significantly) more experiments
 with best--fit masses and cross sections
 larger than the input values.
 Quantitatively,
 for a WIMP mass of 100 GeV with $\sim$ 80 events,
 the 1$\sigma$ and 2$\sigma$ statistical uncertainties
 are $^{+40}_{-35}$ GeV and $^{+100}_{\~-50}$ GeV,
 respectively
 \cite{Green-mchi07}.

\begin{figure}[p!]
\begin{center}
\imageswitch{
\begin{picture}(14,10.25)
\put(0,0){\framebox(14,10.25){Green-mchi}}
\end{picture}}
{
\includegraphics[width=13.5cm]{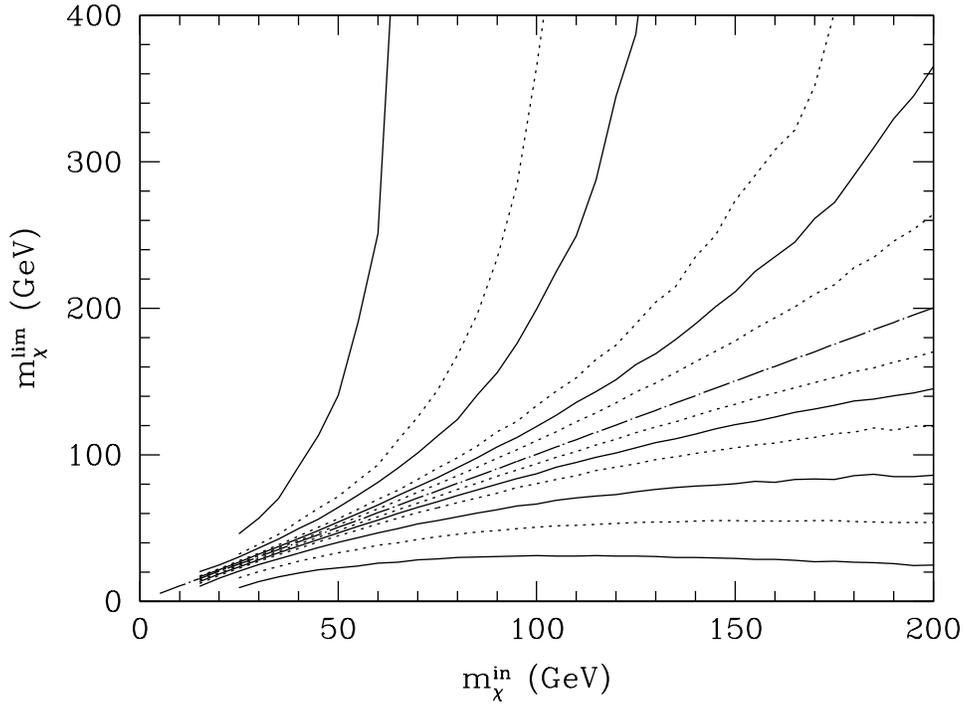}
\vspace*{-0.75cm}
}
\end{center}
\caption{
 The 95\% (solid) and 68\% (dotted) confidence limits
 on the best--fit WIMP mass
 as functions of the input WIMP mass.
 The input SI WIMP--proton cross section
 has been set here as $10^{-8}$ pb.
 The assumed exposures are
 $3 \times 10^3$, $3 \times 10^4$, and $3 \times 10^5$ kg-day,
 respectively
 (Plot from \cite{Green-mchi08}).
}
\label{fig:Green_mchi}
\end{figure}

 Fig.~\ref{fig:Green_mchi} shows
 the 95\% (solid) and 68\% (dotted) confidence limits
 on the best--fit WIMP mass
 as functions of the input WIMP mass.
 The input SI WIMP--proton cross section
 has been set here as $10^{-8}$ pb.
 The assumed exposures are
 $3 \times 10^3$, $3 \times 10^4$, and $3 \times 10^5$ kg-day,
 respectively.
 We see here that since,
 as mentioned above,
 the shape of the recoil spectrum
 varies significantly with the WIMP mass
 for light WIMP masses ($\mchi < \mN$),
 the WIMP mass (and also the cross section)
 can be fitted with a higher accuracy:
 the 1$\sigma$ and 2$\sigma$ statistical uncertainties
 for $\mchi = 25$ GeV
 are $\pm 4$ GeV and $^{+8}_{-7}$ GeV,
 for $\mchi = 50$ GeV
 are $^{+15}_{-12}$ GeV and $^{+22}_{-19}$ GeV,
 respectively
 \cite{Green-mchi07}.

 In contrast,
 the weak dependence of the shape of the recoil spectrum on the WIMP mass
 for heavy WIMP masses ($\mchi \gg \mN$) means that
 it will be more difficult or even impossible
 to extract the WIMP mass
 with $\cal O$(100) events,
 if WIMPs are (much) heavier than the target nucleus
 \cite{Green-mchi07}.
 Note that
 the dependence of the shape of the recoil spectrum
 on the WIMP mass as well as on that of the target nucleus
 suggests that
 heavy nuclei, e.g., Xe,
 would be able to measure the mass of heavy WIMPs more accurately;
 however,
 the rapid decrease of the nuclear form factor
 with increasing recoil energy,
 which occurs for heavy nuclei,
 means that,
 due to less expected events,
 this is in fact not necessarily the case.

\begin{figure}[p!]
\begin{center}
\imageswitch{
\begin{picture}(16,6.4)
\put(0,0){\framebox(16,6.4){Green-sigmapSI-mchi-v0}}
\end{picture}}
{
\includegraphics[width=16cm]{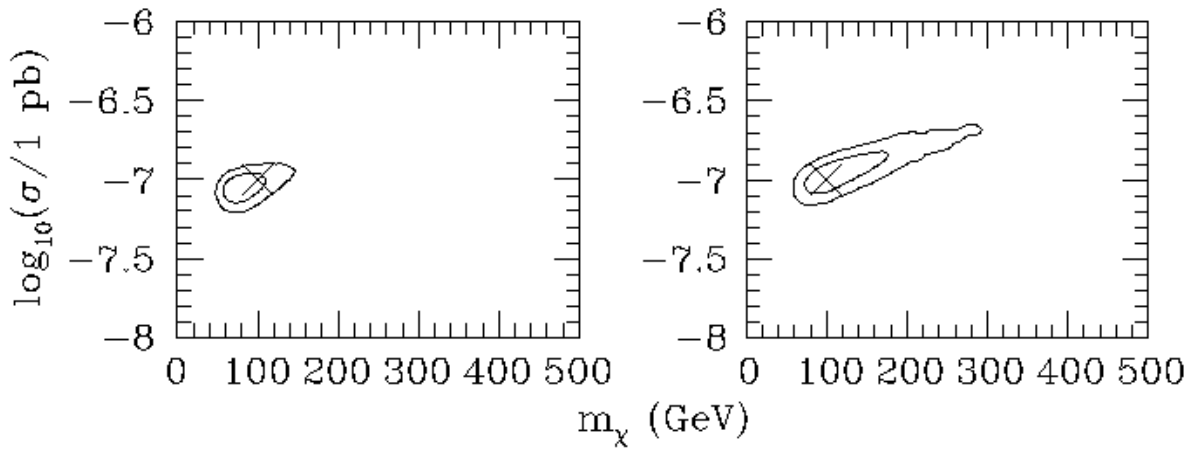}
\vspace*{-0.25cm}
}
\end{center}
\caption{
 Distributions of the best--fit WIMP mass
 and SI WIMP--proton cross section
 on the cross section versus WIMP mass plane.
 The input orbital velocity of the Solar system
 $v_0$ has been set as 200 (left) and 240 (right) km/s,
 while the standard value of $v_0 = 220$ km/s has been used
 for the data analysis.
 The exposure assumed here is $3 \times 10^3$ kg-day.
 The other parameters are as in Figs.~\ref{fig:Green_sigmapSI_mchi}
 (Plots from \cite{Green-mchi07}).
}
\label{fig:Green_sigmapSI_mchi_v0}
\end{figure}
\subsubsection{Systematic uncertainties}
 Different sources of the systematic uncertainties
 in this model--dependent analysis
 have been considered \cite{{Green-mchi07}, {Green-mchi08}}.
 Figs.~\ref{fig:Green_sigmapSI_mchi_v0} show
 the distributions of the best--fit WIMP mass and cross section
 with different {\em input} orbital velocity of the Solar system:
 $v_0 = 200$ (left) and 240 (right) km/s,
 while the standard value of $v_0 = 220$ km/s
 has been used for the data analysis.
 As shown here,
 for an input WIMP mass of 100 GeV,
 there could be an $\sim$ $\pm$20 GeV shift in the best--fit WIMP mass
 combined with an $\sim \pm 10^{-8}$ pb ($\sim$ 10\%) shift
 in the SI WIMP--proton cross section
 caused by the $\pm$ 20 km/s difference
 between the real and the assumed orbital velocities
 \cite{Green-mchi07}.
 Moreover,
 the larger the real orbital velocity,
 the less the expected event number (with a fixed exposure),
 and thus the larger the statistical uncertainties
 on both the WIMP mass and SI WIMP--proton cross section
 one could obtain.

 More detailed illustrations and discussions about
 the effects of varying the underlying WIMP mass and cross section,
 the detector target nucleus,
 the exposure, the minimal and maximal cut--off energies,
 the orbital velocity of the Solar system,
 as well as the background event rate and its spectrum
 can be found in Refs.~\cite{{Green-mchi07}, {Green-mchi08}, {Bernal08}}.
\section{Combining two experiments}
 In this section
 I first review the model--independent method
 for reconstructing the WIMP mass
 by using two experimental data sets
 with different target nuclei%
\footnote{
 In Ref.~\cite{Jackson06},
 the authors mentioned an attempt
 for using the maximum likelihood analysis
 with two (or more) detector materials.
 However,
 they found that,
 since the likelihood contours for different targets
 are pretty similar
 when simulating with the same number of events,
 their results showed effectively little different
 from that obtained with a single experiment.
}.
 Then I also describe an extension of this method
 for estimating (or at least constraining)
 the SI WIMP--proton cross section.
\subsection{Model--independent determination}
 As mentioned in the introduction,
 the normalized one--dimensional velocity distribution function
 of incident WIMPs can be solved
 from Eq.(\ref{eqn:dRdQ}) directly and,
 consequently,
 its generalized moments can be estimated by \cite{DMDDmchi}
\beqn
    \expv{v^n}(v(\Qmin), v(\Qmax))
 \= \int_{v(\Qmin)}^{v(\Qmax)} v^n f_1(v) \~ dv
    \non\\
 \= \alpha^n
    \bfrac{2 \Qmin^{(n+1)/2} r(\Qmin) / \FQmin + (n+1) I_n(\Qmin, \Qmax)}
          {2 \Qmin^{   1 /2} r(\Qmin) / \FQmin +       I_0(\Qmin, \Qmax)}
\~.
\label{eqn:moments}
\eeqn
 Here $v(Q) = \alpha \sqrt{Q}$,
 $Q_{\rm (min, max)}$ are the experimental minimal and maximal cut--off energies,
\beq
        r(\Qmin)
 \equiv \adRdQ_{{\rm expt},\~Q = \Qmin}
\label{eqn:rmin}
\eeq
 is an estimated value of the {\em measured} recoil spectrum
 $(dR/dQ)_{\rm expt}$ ({\em before} the normalization by the exposure $\cal E$) at $Q = \Qmin$,
 and $I_n(\Qmin, \Qmax)$ can be estimated through the sum:
\beq
   I_n(\Qmin, \Qmax)
 = \sum_a \frac{Q_a^{(n-1)/2}}{F^2(Q_a)}
\~,
\label{eqn:In_sum}
\eeq
 where the sum runs over all events in the data set
 that satisfy $Q_a \in [\Qmin, \Qmax]$.
 Note that,
 firstly,
 by using the second Eq.(\ref{eqn:moments})
 $\expv{v^n}(v(\Qmin), v(\Qmax))$ can be determined
 independently of the local WIMP density $\rho_0$,
 of the velocity distribution function of incident WIMPs, $f_1(v)$,
 as well as of the WIMP--nucleus cross section $\sigma_0$.
 Secondly,
 as shown later,
 $r(\Qmin)$ and $I_n(\Qmin, \Qmax)$
 are two key quantities for this model--independent method,
 which can be estimated
 either from a functional form of the recoil spectrum
 or from experimental data (i.e., the measured recoil energies) directly%
\footnote{
 All formulae needed for estimating
 $r(\Qmin)$, $I_n(\Qmin, \Qmax)$, and their statistical errors
 are given in the appendix.
}.
 However,
 $r(\Qmin)$ and $I_n(\Qmin, \Qmax)$ estimated
 from a scattering spectrum fitted to experimental data
 are not model--independent any more.
\subsubsection{Basic expressions for determining \boldmath$\mchi$}
 By requiring that
 the values of a given moment of $f_1(v)$
 estimated by Eq.(\ref{eqn:moments})
 from two detectors with different target nuclei, $X$ and $Y$, agree,
 $\mchi$ appearing in the prefactor $\alpha^n$
 on the right--hand side of Eq.(\ref{eqn:moments})
 can be solved as
 \cite{DMDDmchi-SUSY07}:
\beq
   \left. \mchi \right|_{\Expv{v^n}}
 = \frac{\sqrt{\mX \mY} - \mX (\calR_{n, X} / \calR_{n, Y})}
        {\calR_{n, X} / \calR_{n, Y} - \sqrt{\mX / \mY}}
\~,
\label{eqn:mchi_Rn}
\eeq
 where
\beqn
        \calR_{n, X}
 \equiv \bfrac{2 \QminX^{(n+1)/2} r_X(\QminX) / \FQminX + (n+1) \InX}
              {2 \QminX^{   1 /2} r_X(\QminX) / \FQminX +       \IzX}^{1/n}
\~,
\label{eqn:Rn_min}
\eeqn
 and $\calR_{n, Y}$ can be defined analogously.
 Here $n \ne 0$,
 $m_{(X, Y)}$ and $F_{(X, Y)}(Q)$
 are the masses and the form factors of the nucleus $X$ and $Y$,
 respectively,
 and $r_{(X, Y)}(Q_{{\rm min}, (X, Y)})$
 refer to the counting rates for detectors $X$ and $Y$
 at the respective lowest recoil energies included in the analysis.
 Note that,
 firstly,
 the general expression (\ref{eqn:mchi_Rn}) can be used
 either for spin--independent or for spin--dependent scattering,
 one only needs to choose different form factors under different assumptions.
 Secondly,
 the form factors in the estimate of $\InX$ and $\InY$
 using Eq.(\ref{eqn:In_sum}) are also different.

 On the other hand,
 by using the theoretical prediction that
 the SI WIMP--nucleus cross section
 given in Eq.(\ref{eqn:sigma0SI}) dominates,
 and the fact that
 the integral over the one--dimensional WIMP velocity distribution
 on the right--hand side of Eq.(\ref{eqn:dRdQ})
 is the minus--first moment of this distribution,
 which can be estimated by Eq.(\ref{eqn:moments}) with $n = -1$,
 one can easily find that \cite{DMDDmchi}
\beq
   \rho_0 |f_{\rm p}|^2
 = \frac{\pi}{4 \sqrt{2}} \afrac{\mchi + \mN}{\calE A^2 \sqrt{\mN}}
   \bbrac{\frac{2 \Qmin^{1/2} r(\Qmin)}{\FQmin} + I_0}
\~.
\label{eqn:rho0_fp2}
\eeq
 Note that
 the exposure of the experiment, $\calE$,
 appears in the denominator.
 Since the unknown factor $\rho_0 |f_{\rm p}|^2$ on the left--hand side above
 is identical for different targets,
 it leads to a second expression for determining $\mchi$
 \cite{DMDDmchi}
\beq
   \left. \mchi \right|_\sigma
 = \frac{\abrac{\mX / \mY}^{5/2} \mY - \mX (\calR_{\sigma, X} / \calR_{\sigma, Y})}
        {\calR_{\sigma, X} / \calR_{\sigma, Y} - \abrac{\mX / \mY}^{5/2}}
\~.
\label{eqn:mchi_Rsigma}
\eeq
 Here $m_{(X, Y)} \propto A_{(X, Y)}$ has been assumed,
\beq
        \calR_{\sigma, X}
 \equiv \frac{1}{\calE_X}
        \bbrac{\frac{2 \QminX^{1/2} r_X(\QminX)}{\FQminX} + \IzX}
\~,
\label{eqn:Rsigma_min}
\eeq
 and similarly for $\calR_{\sigma, Y}$.
\subsubsection{\boldmath$\chi^2$--fitting}
 In order to yield the best--fit WIMP mass
 as well as to minimize its statistical error
 by combining the estimators for different $n$
 in Eq.(\ref{eqn:mchi_Rn}) with each other
 and with the estimator in Eq.(\ref{eqn:mchi_Rsigma}),
 a $\chi^2$ function has been introduced \cite{DMDDmchi}
\beq 
   \chi^2(\mchi)
 = \sum_{i, j}
   \abrac{f_{i, X} - f_{i, Y}} {\cal C}^{-1}_{ij} \abrac{f_{j, X} - f_{j, Y}}
\~,
\label{eqn:chi2}
\eeq
 where
\cheqna
\beqn
           f_{i, X}
 \eqnequiv \alpha_X^i
           \bfrac{  2 Q_{{\rm min}, X}^{(i+1)/2} r_X(\Qmin) / F_X^2(Q_{{\rm min}, X})
                  + (i+1) I_{i,X}}
                 {  2 Q_{{\rm min}, X}^{   1 /2} r_X(\Qmin) / F_X^2(Q_{{\rm min}, X})
                  +       \IzX}
           \afrac{1}{300~{\rm km/s}}^i
           \non\\
           \non\\
 \=        \afrac{\alpha_X {\cal R}_{i, X}}{300~{\rm km/s}}^{i}
\~,
\label{eqn:fiXa}
\eeqn
 for $i = -1,~1,~2,~\dots,~n_{\rm max}$, and
\cheqnb
\beqn
           f_{n_{\rm max}+1, X}
 \eqnequiv \calE_X
           \bfrac{A_X^2}
                 {  2 Q_{{\rm min}, X}^{1/2} r_X(\Qmin) / F_X^2(Q_{{\rm min}, X})
                  + \IzX}
           \afrac{\sqrt{\mX}}{\mchi + \mX}
           \non\\
           \non\\
 \=        \frac{A_X^2}{\calR_{\sigma, X}} \afrac{\sqrt{\mX}}{\mchi + \mX}
\~;
\label{eqn:fiXb}
\eeqn
\cheqn
 the other $n_{\rm max} + 2$ functions $f_{i,Y}$ can be defined analogously.
 Here $n_{\rm max}$ determines the highest moment of $f_1(v)$
 that is included in the fit.
 The $f_i$ are normalized such that
 they are dimensionless and very roughly of order unity
 in order to alleviate numerical problems
 associated with the inversion of their covariance matrix.
 Note that
 the first $n_{\rm max} + 1$ fit functions
 depend on $\mchi$ only through the overall factor $\alpha$
 and that
 $\mchi$ in Eqs.(\ref{eqn:fiXa}) and (\ref{eqn:fiXb})
 is now a fit parameter,
 which may differ from the true value of the WIMP mass.
 Finally,
 $\cal C$ in Eq.(\ref{eqn:chi2}) is the total covariance matrix.
 Since the $X$ and $Y$ quantities
 are statistically completely independent,
 $\cal C$ can be written as a sum of two terms%
\footnote{
 Formulae needed for estimating the entries of $\cal C$
 will be given in the appendix.
}:
\beq
   {\cal C}_{ij}
 = {\rm cov}\abrac{f_{i, X}, f_{j, X}} + {\rm cov}\abrac{f_{i, Y}, f_{j, Y}}
\~.
\label{eqn:Cij}
\eeq
\subsubsection{Matching the cut--off energies}
 The basic requirement of the expressions for determining $\mchi$
 given in Eqs.(\ref{eqn:mchi_Rn}) and (\ref{eqn:mchi_Rsigma}) is that,
 from two experiments with different target nuclei,
 the values of a given moment of the WIMP velocity distribution
 estimated by Eq.(\ref{eqn:moments}) should agree.
 This means that
 the upper cuts on $f_1(v)$ in two data sets
 should be (approximately) equal%
\footnote{
 Here the threshold energies have been assumed to be negligibly small.
}.
 Since $v_{\rm cut} = \alpha \sqrt{Q_{\rm max}}$,
 it requires that \cite{DMDDmchi}
\beq
   Q_{{\rm max}, Y}
 = \afrac{\alpha_X}{\alpha_Y}^2 Q_{{\rm max}, X}
\~.
\label{eqn:match}  
\eeq
 Note that
 $\alpha$ defined in Eq.(\ref{eqn:alpha})
 is a function of the true WIMP mass.
 Thus this relation for matching optimal cut--off energies
 can be used only if $\mchi$ is already known.
 One possibility to overcome this problem is
 to fix the cut--off energy of the experiment with the heavier target,
 minimize the $\chi^2(\mchi)$ function
 defined in Eq.(\ref{eqn:chi2}),
 and estimate the cut--off energy for the lighter nucleus
 by Eq.(\ref{eqn:match}) algorithmically \cite{DMDDmchi}.
\subsection{Numerical results}
 Here I show some numerical results
 for the reconstructed WIMP mass
 based on Monte Carlo simulations.
 The upper and lower bounds on the reconstructed WIMP mass
 are estimated from the requirement that
 $\chi^2$ exceeds its minimum by 1%
\footnote{
 The median, rather than the mean, values
 for the (bounds on the) reconstructed WIMP mass are shown.
}.
 $\rmXA{Si}{28}$ and $\rmXA{Ge}{76}$
 have been chosen as two target nuclei.
 The scattering cross section has been assumed
 to be dominated by spin--independent interactions.
 The shifted Maxwellian velocity distribution
 given in Eq.(\ref{eqn:f1v_sh_ast})
 (the second term involving $\vesc$ has been neglected)
 with $v_0 = 220$ km/s, $v_{\rm e} = 1.05 \~ v_0$%
\footnote{
 The time dependence of the Earth's velocity
 in the Galactic frame,
 the second term of $v_{\rm e}(t)$ in Eq.(\ref{eqn:ve}),
 has been ignored.
},
 and \mbox{$\vesc = 700$ km/s}
 and the Woods-Saxon form factor in Eq.(\ref{eqn:FQ_SI_WS})
 have been used.
 The threshold energies of two experiments
 have been assumed to be negligible
 and the maximal experimental cut--off energies are set as 100 keV.
 2 $\times$ 5,000 experiments have been simulated.
 In order to avoid large contributions
 from very few events in the high energy range
 to the higher moments \cite{DMDDf1v},
 only the moments up to $n_{\rm max} = 2$
 were included in the $\chi^2$ fit.
\subsubsection{Statistical uncertainty}
\begin{figure}[p!]
\begin{center}
\imageswitch{
\begin{picture}(12,22)
\put(0,12){\framebox(12,10){m50-100-sige}}
\put(0, 0){\framebox(12,10){m500-sige}}
\end{picture}}
{
\vspace*{-1cm}
\rotatebox{-90}{\includegraphics[width=12cm]{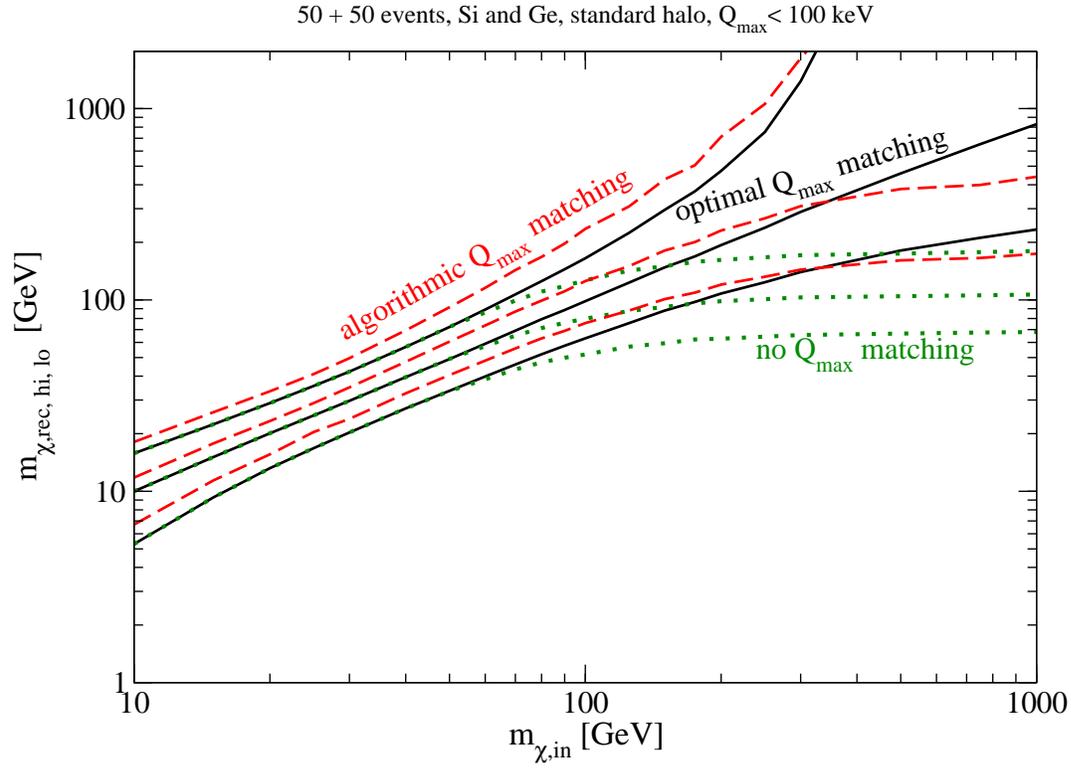}}
\rotatebox{-90}{\includegraphics[width=12cm]{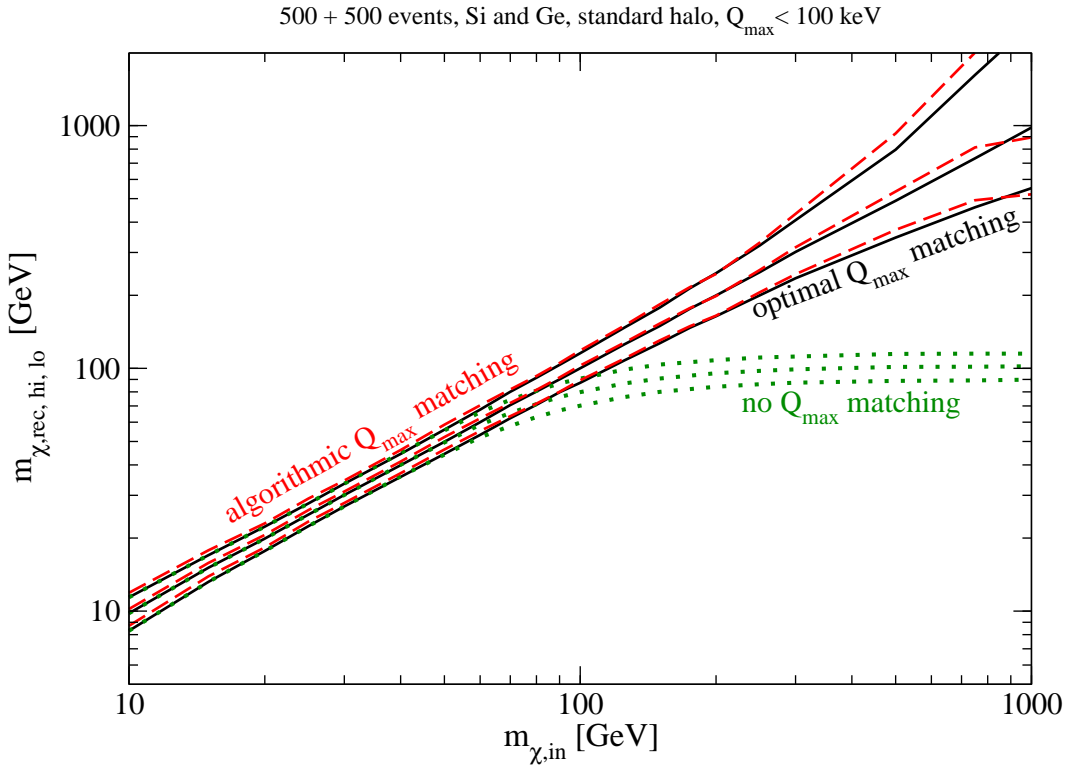}}
\vspace{-1cm}
}
\end{center}
\caption{
 Results for the reconstructed WIMP mass
 as well as its $1 \sigma$ statistical error interval
 based on the $\chi^2$--fit in Eq.(\ref{eqn:chi2}).
 50 (upper) and 500 (lower) events on average before cuts
 from each experiment have been simulated.
 See the text for further details
 (Plots from Ref.~\cite{DMDDmchi}).
}
\label{fig:mchi_rec}
\end{figure}
 In Figs.~\ref{fig:mchi_rec}
 the dotted (green) curves show the median reconstructed WIMP mass
 and its $1 \sigma$ upper and lower bounds
 for the case that
 both $Q_{\rm max, Si} $ and $Q_{\rm max, Ge}$ have been fixed to 100 keV.
 As argued earlier,
 the values of a given moment of the WIMP velocity distribution
 estimated by Eq.(\ref{eqn:moments})
 do not agree
 when the same maximal cut--off energy
 for both experimental data sets is used.
 This causes a systematic {\em underestimate}
 of the reconstructed WIMP mass \cite{DMDDmchi-SUSY07}
 which can be seen obviously here.

 The solid (black) curves
 were obtained by using Eq.(\ref{eqn:match})
 for matching the cut--off energy $Q_{\rm max, Si}$ perfectly
 with $Q_{\rm max, Ge} = 100$ keV
 and the true (input) WIMP mass,
 whereas
 the dashed (red) curves show the case
 that $Q_{\rm max, Ge} = 100$ keV,
 and $Q_{\rm max, Si}$ has been determined
 by minimizing $\chi^2(\mchi;Q_{\rm max, Si})$.
 As shown here,
 with only 50 events on average before cuts
 (upper frame) from each experiment,
 the algorithmic process seems already to work pretty well
 for WIMP masses up to $\sim 500$ GeV.
 For $\mchi \lsim 100$ GeV
 the median WIMP mass determined in this way
 {\em overestimates} its true value by 15 to 20\%;
 however,
 the true WIMP mass always lies
 within the median limits of the $1 \sigma$ statistical error interval
 estimated by the algorithmic $\Qmax$ matching procedure
 up to even $\mchi = 1$ TeV \cite{DMDDmchi}.
\subsubsection{Statistical fluctuation}
\begin{figure}[t!]
\begin{center}
\imageswitch{
\begin{picture}(12,9)
\put(0,0){\framebox(12,9){del50-50-sige}}
\end{picture}}
{
\vspace*{-1cm}
\rotatebox{-90}{\includegraphics[width=12cm]{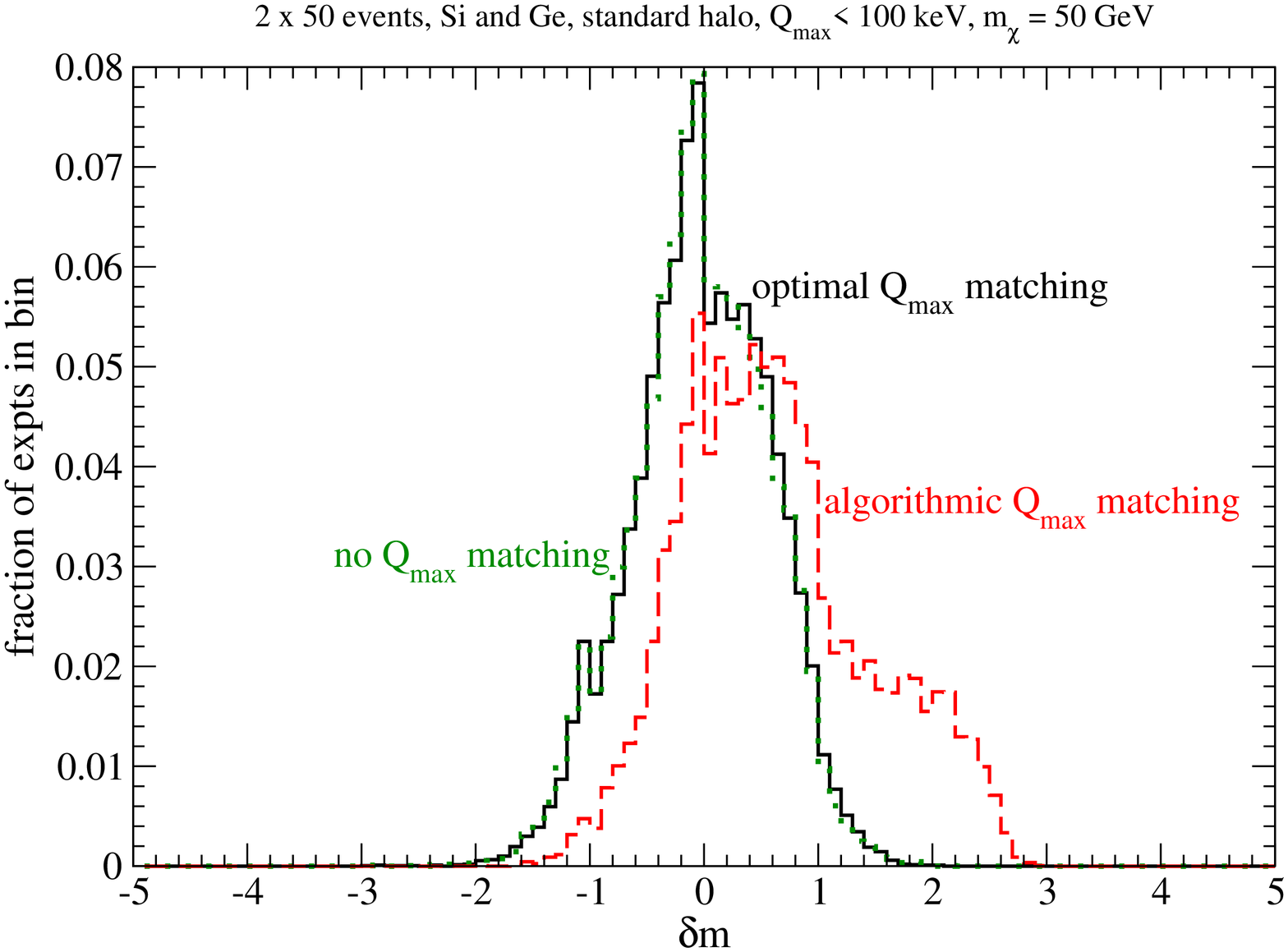}}
\vspace*{-1cm}
}
\end{center}
\caption{
 Normalized distribution of the estimator $\delta m$
 defined in Eq.(\ref{eqn:deltam})
 for an input WIMP mass of 50 GeV
 with 50 events on average (before cuts)
 in each experiment.
 The other parameters and notations are as in Figs.~\ref{fig:mchi_rec}
 (Plot from Ref.~\cite{DMDDmchi}).
}
\label{fig:del50}
\end{figure}

 In order to study
 the statistical fluctuation of the reconstructed WIMP mass
 by algorithmic $\Qmax$ matching in the simulated experiments,
 an estimator $\delta m$ has been introduced as \cite{DMDDmchi}
\beq
\renewcommand{\arraystretch}{0.75}
   \delta m
 = \left\{
    \begin{array}{l c l}
     \D   1
        + \frac{m_{\chi, {\rm lo1}} - m_{\chi, {\rm in }}}
               {m_{\chi, {\rm lo1}} - m_{\chi, {\rm lo2}}}\~, & ~~~~~~ & 
        {\rm if}~m_{\chi, {\rm in }} \leq m_{\chi, {\rm lo1}}\~;
     \\ & & \\
    \D   \frac {m_{\chi, {\rm rec}} - m_{\chi, {\rm in }}}
               {m_{\chi, {\rm rec}} - m_{\chi, {\rm lo1}}}\~, &        &
        {\rm if}~m_{\chi,{\rm lo1}} < m_{\chi, {\rm in }} < m_{\chi, {\rm rec}}\~;
    \\ & & \\
    \D   \frac {m_{\chi, {\rm rec}} - m_{\chi, {\rm in }}}
               {m_{\chi, {\rm hi1}} - m_{\chi, {\rm rec}}}\~, &        &
        {\rm if}~m_{\chi,{\rm rec}} < m_{\chi, {\rm in }} < m_{\chi, {\rm hi1}}\~;
    \\ & & \\
    \D   \frac {m_{\chi, {\rm hi1}} - m_{\chi, {\rm in }}}
               {m_{\chi, {\rm hi2}} - m_{\chi, {\rm hi1}}} - 1 \~, &   &
        {\rm if}~m_{\chi, {\rm in }} \geq m_{\chi,{\rm hi1}}\~.
    \end{array}
   \right.
\label{eqn:deltam}
\eeq
 Here $m_{\chi, {\rm in }}$ is the true (input) WIMP mass,
 $m_{\chi, {\rm rec}}$ its reconstructed value,
 $m_{\chi, {\rm lo1(2)}}$ are the $1 \~ (2) \~ \sigma$ lower bounds
 satisfying $\chi^2(m_{\chi, {\rm lo(1, 2)}}) = \chi^2(m_{\chi, {\rm rec}}) + 1 \~ (4)$,
 and $m_{\chi, {\rm hi1(2)}}$ are
 the corresponding $1 \~ (2) \~ \sigma$ upper bounds.
 It has been found that
 the error intervals of the median reconstructed WIMP mass
 are quite asymmetric;
 similarly,
 the distance between the $2 \sigma$ and $1 \sigma$ limits
 can be quite different from
 the distance between the $1 \sigma$ limit and the central value
 \cite{DMDDmchi}%
\footnote{
 Recall that
 the same asymmetry has also been observed
 by the maximum likelihood analysis.
}.
 The definition of $\delta m$ in Eq.(\ref{eqn:deltam})
 takes these differences into account,
 and also keeps track of the sign of the deviation:
 if the reconstructed WIMP mass is larger (smaller) than the true one,
 $\delta m$ is positive (negative).
 Moreover,
 $|\delta m| \leq 1 \~ (2)$ if and only if the true WIMP mass lies between the
 experimental $1 \~ (2) \~ \sigma$ limits.

 Fig.~\ref{fig:del50} shows the distribution of $\delta m$
 calculated from 5,000 simulated experiments
 with 50 events on average before cuts
 for a rather light WIMP mass of 50 GeV.
 In this case
 simply fixing both $\Qmax$ values to 100 keV still works fine
 (see the upper frame of Figs.~\ref{fig:mchi_rec}).
 However,
 the distributions for both fixed $\Qmax$ and optimal $\Qmax$ matching
 look somewhat lopsided,
 since the error interval is already asymmetric,
 with $  m_{\chi, {\rm hi1}} - m_{\chi, {\rm rec}}
       > m_{\chi, {\rm rec}} - m_{\chi, {\rm lo1}}$.
 The overestimate of light WIMP masses
 reconstructed by algorithmic $\Qmax$ matching
 shown in Figs.~\ref{fig:mchi_rec}
 is reflected by the dashed (red) histogram here,
 which has significantly more entries
 at positive values than at negative values.
 These distributions also indicate that
 the statistical uncertainties estimated
 by minimizing $\chi^2(\mchi)$ are indeed overestimated,
 since nearly 90\% of the simulated experiments
 have $|\delta m| \leq 1$ \cite{DMDDmchi},
 much more than $\sim$ 68\% of the experiments,
 that a usual $1 \sigma$ error interval should contain.

\begin{figure}[t!]
\begin{center}
\imageswitch{
\vspace*{0.5cm}
\begin{picture}(16.5,8)
\put(0  ,0){\framebox(8,8){del50-200-sige}}
\put(8.5,0){\framebox(8,8){del500-200-sige}}
\end{picture}}
{
\rotatebox{-90}{\includegraphics[width=8.5cm]{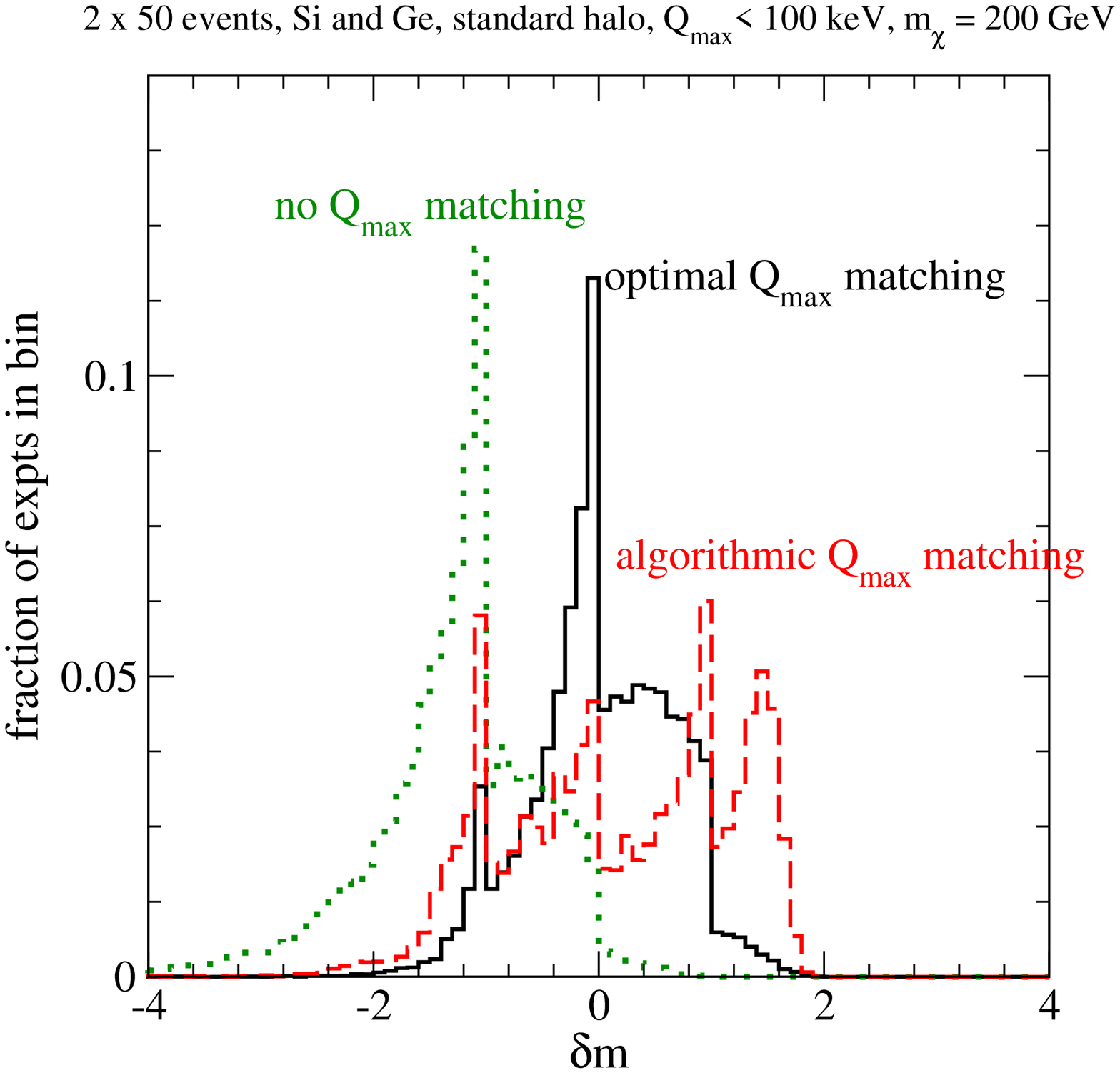}} 
\rotatebox{-90}{\includegraphics[width=8.5cm]{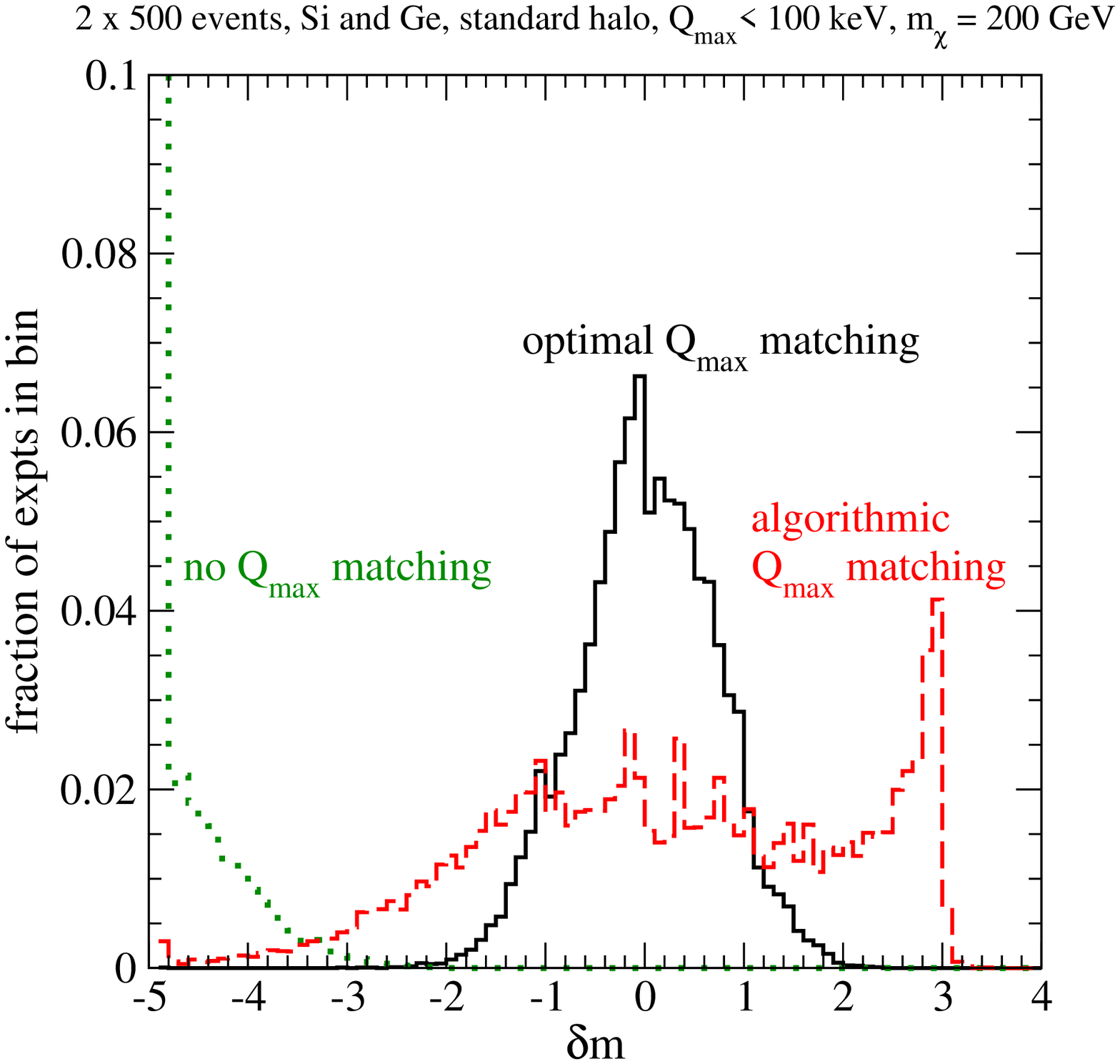}} \\
\vspace*{-1cm}
}
\end{center}
\caption{
 Normalized distribution of the estimator $\delta m$.
 Parameters and notations are as in Fig.~\ref{fig:del50}, 
 except that the input WIMP mass has been increased to 200 GeV.
 In the right frame
 the average event number (before cuts) in each experiment
 have, in addition, been increased from 50 to 500.
 Note that
 the bins at $\delta m = \pm 5$ are overflow bins,
 i.e., they also contain all experiments with $|\delta m| \ge 5$.
 (Plots from Ref.~\cite{DMDDmchi}).
}
\label{fig:del200}
\end{figure}

 Unfortunately,
 as shown in Figs.~\ref{fig:del200},
 when the true (input) WIMP mass increases to 200 GeV
 and the expected event number (before cuts)
 increases to 500 (right frame),
 the situations become less favorable.
 While
 optimal $\Qmax$ matching seems to approach very slowly
 to be Gaussian and
 the overestimated statistical errors
 become a little bit more reliable for larger event numbers
 \cite{DMDDmchi},
 the errors estimated by
 the algorithmic procedure for determining $Q_{\rm max, Si}$
 are not very reliable in the simulations.

 More detailed illustrations and discussions
 about algorithmic $\Qmax$ matching
 with different detector materials
 or with data sets generated in different halo models,
 as well as about the statistical fluctuation
 in the analysis can be found in Ref.~\cite{DMDDmchi}.
\subsection{Estimating the SI WIMP--proton coupling}
 In the maximum likelihood analysis discussed in Sec.~2,
 the SI WIMP--proton cross section $\sigmapSI$
 is the second fitting parameter that,
 combined with the WIMP mass $\mchi$, 
 maximizes the extended likelihood function $\cal L$
 calculated from an assumed WIMP velocity distribution.

 In contrast,
 as shown above,
 by combining two experimental data sets,
 one can estimate the WIMP mass $\mchi$
 without knowing the WIMP--nucleus cross section $\sigma_0$.
 Conversely,
 by means of Eq.(\ref{eqn:rho0_fp2}),
 one can also estimate or at least constrain
 the SI WIMP--proton coupling, $|f_{\rm p}|^2$,
 from experimental data directly
 without knowing the WIMP mass
 \cite{DMDDfp2-IDM2008}.
\subsubsection{Making an assumption for the local WIMP density}
 In Eq.(\ref{eqn:rho0_fp2})
 the WIMP mass $\mchi$ on the right--hand side
 can be determined by the method described above,
 $r(\Qmin)$ and $I_0$ can also be estimated
 from one of the two data sets used for determining $\mchi$
 or from a third experiment.
 Nevertheless,
 due to the degeneracy between
 the local WIMP density $\rho_0$
 and the coupling $|f_{\rm p}|^2$,
 one {\em cannot} estimate both of them independently.
 The simplest way is making
 an assumption for the local WIMP density $\rho_0$%
\footnote{
 Note that,
 since the coupling $|f_{\rm p}|^2$
 estimated  by Eq.(\ref{eqn:rho0_fp2}) 
 is inversely proportional to the local density $\rho_0$,
 whose common value falls on the lower end of the possible range
 (see Eqs.(\ref{eqn:rho0}) and (\ref{eqn:rho0_range})),
 one can therefore at least give an {\em upper bound} on this coupling.
}.
\subsubsection{Numerical results}
\begin{figure}[t]
\begin{center}
\imageswitch{
\begin{picture}(16,7.5)
\put(0  ,0){\framebox(7.5,7.5){fp-mchi-0500-050-Ge}}
\put(8.5,0){\framebox(7.5,7.5){fp2-mchi-sige-Ge-050}}
\end{picture}}
{
\includegraphics[width=7.5cm]{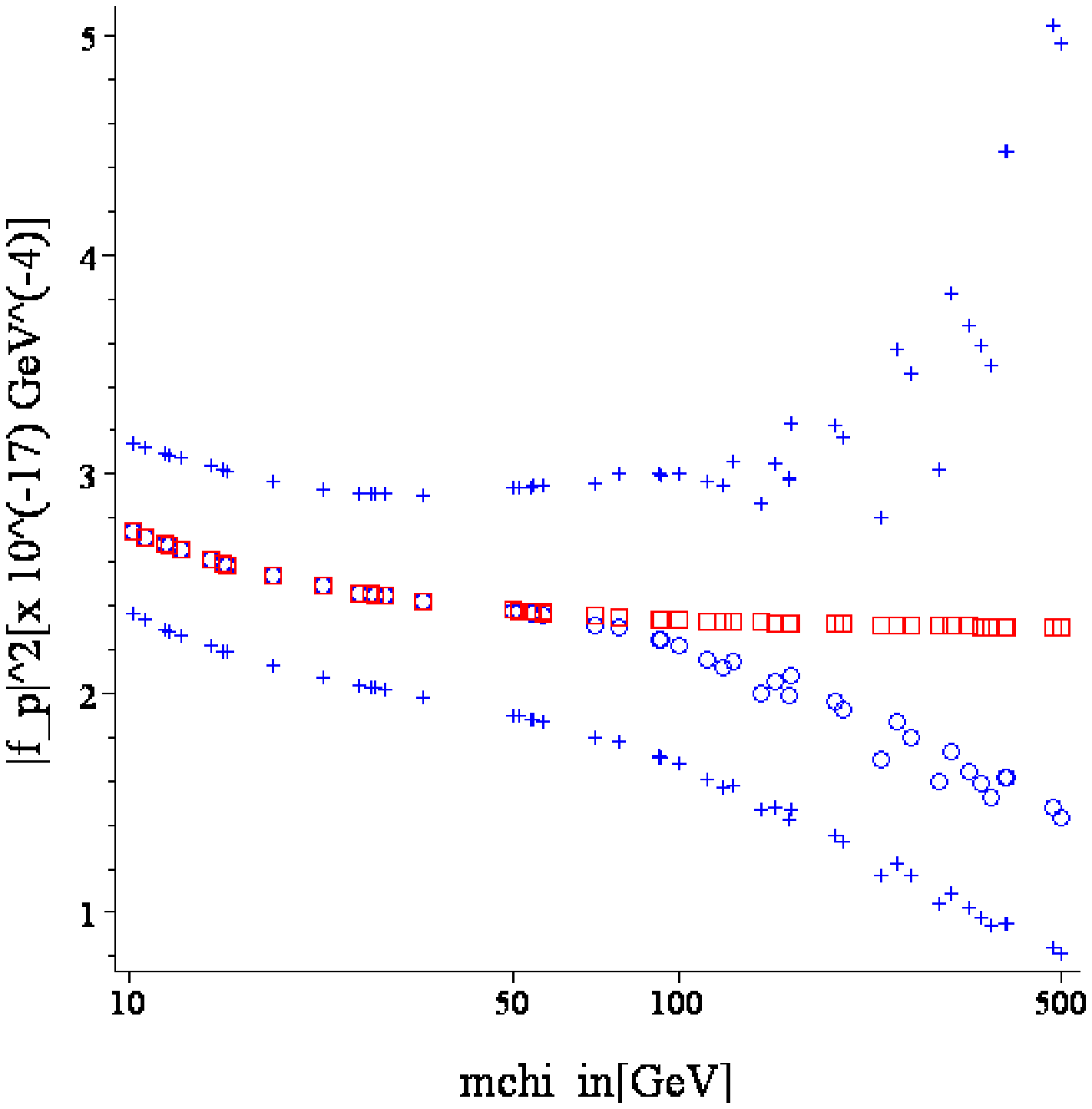}      \hspace{1cm}
\includegraphics[width=7.5cm]{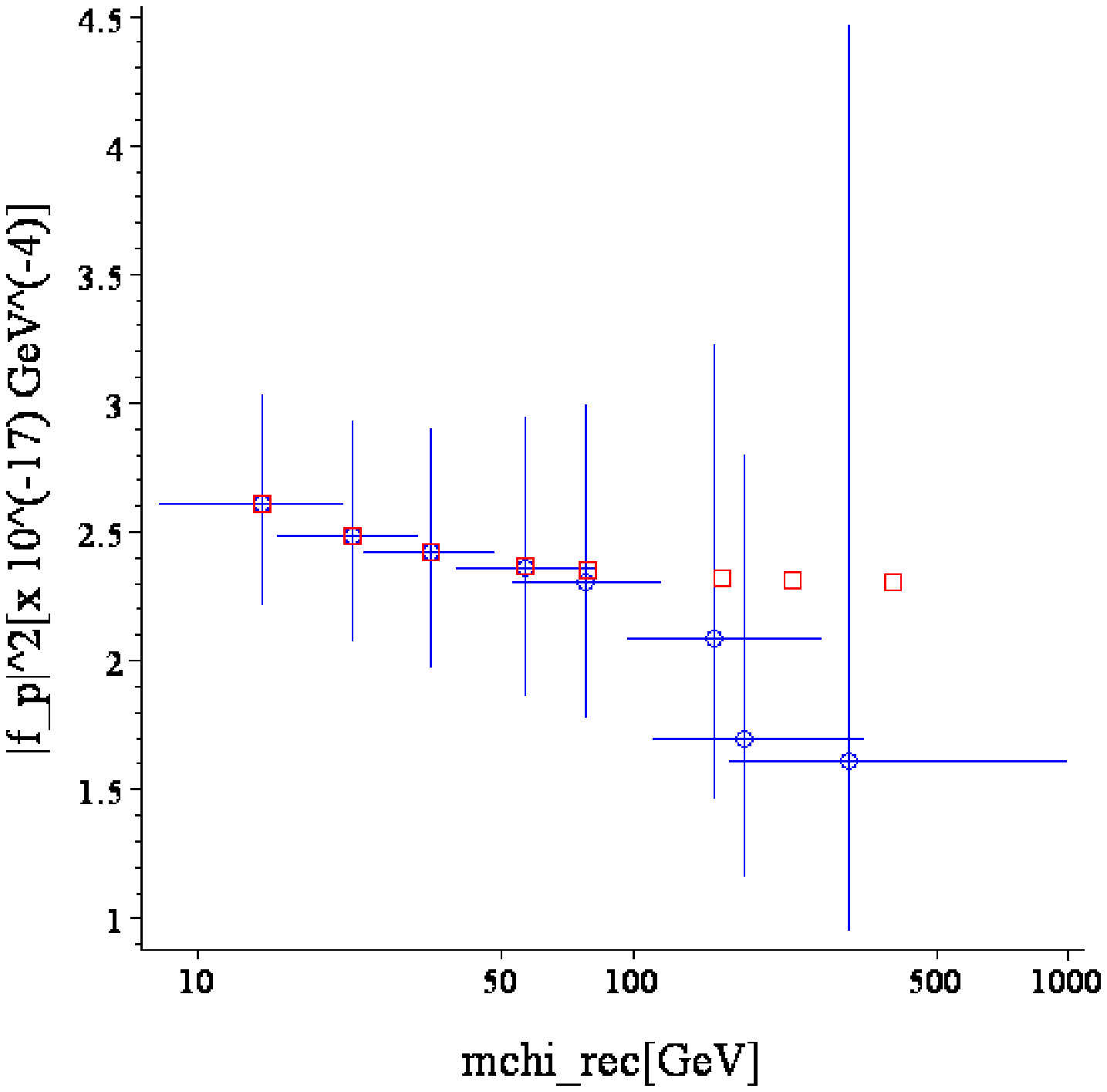}
\vspace*{-0.5cm}
}
\end{center}
\caption{
 Left:
 The reconstructed SI WIMP--proton coupling $|f_{\rm p}|^2_{\rm rec}$
 as a function of the {\em input} WIMP mass $m_{\chi, {\rm in}}$.
 Right:
 The reconstructed coupling $|f_{\rm p}|^2_{\rm rec}$
 and the {\em reconstructed} WIMP mass $m_{\chi, {\rm rec}}$
 on the cross section (coupling) versus WIMP mass plane.
 The open (red) squares indicate
 the input WIMP masses and
 the true values of the coupling.
 The open (blue) circles and the (blue) crosses indicate
 the reconstructed couplings and
 their $1 \sigma$ statistical errors.
 The horizontal and vertical solid (blue) lines show
 the $1 \sigma$ statistical errors on
 $m_{\chi, {\rm rec}}$ and $|f_{\rm p}|^2_{\rm rec}$,
 respectively.
 Parameters are as in Figs.~\ref{fig:mchi_rec},
 in addition $\sigmapSI$ has been set as $10^{-8}$ pb.
 Each experiment has 50 events on average.
 See the text for further details
 (Plots from Ref.~\cite{DMDDfp2-IDM2008}).
}
\label{fig:fp2_rec}
\end{figure}

 The left frame of Figs.~\ref{fig:fp2_rec}
 shows the reconstructed SI WIMP--proton coupling $|f_{\rm p}|^2_{\rm rec}$
 as a function of the input WIMP mass $m_{\chi, {\rm in}}$.
 Following simulations for the reconstruction of the WIMP mass,
 $\rmXA{Si}{28}$ and $\rmXA{Ge}{76}$ were chosen as two target nuclei
 for estimating $\mchi$ in Eq.(\ref{eqn:rho0_fp2}).
 In order to avoid
 complicated calculations of the correlation
 between the error on the reconstructed $\mchi$
 and that on the estimator of $I_0$,
 a second, independent data set with $\rmXA{Ge}{76}$
 was chosen as the third target
 for estimating $I_0$.
 The SI WIMP--proton cross section was set as $10^{-8}$ pb.
 Each experimental data set
 has 50 events on average
 under the common experimental cut--off energy $\Qmax$ chosen as 100 GeV.

 It can be seen that
 the reconstructed $|f_{\rm p}|^2$ values
 are {\em underestimated} for WIMP masses $\gsim~100$ GeV.
 This systematic deviation is caused mainly
 by the underestimate of $I_0$.
 However,
 in spite of this systematic deviation
 (and in fact due to the fairly large statistical uncertainty),
 the true value of $|f_{\rm p}|^2$
 always lies within the $1 \sigma$ statistical error interval.
 Moreover,
 for a WIMP mass of 100 GeV,
 one could in principle already estimate the SI WIMP--proton coupling
 with a statistical uncertainty of only $\sim$ 15\%
 with just 50 events from each experiment.
 Recall that
 this is much smaller than
 the systematic uncertainty of the local Dark Matter density
 (of a factor of 2 or even larger).

 Combining the estimate for the SI WIMP--proton coupling
 with the estimate for the WIMP mass,
 the right frame of Figs.~\ref{fig:fp2_rec} shows
 the reconstructed coupling $|f_{\rm p}|^2_{\rm rec}$ and
 the {\em reconstructed} WIMP mass $m_{\chi, {\rm rec}}$
 on the cross section (coupling) versus WIMP mass plane%
\footnote{
 Plots shown here have been calculated by a different program
 than that for the Monte Carlo simulations
 shown in Figs.~\ref{fig:mchi_rec} to \ref{fig:del200}.
}.
 It is important to note that,
 as shown here,
 $|f_{\rm p}|^2$ and $\mchi$ can be estimated {\em separately}
 and from experimental data {\em directly} with
 {\em neither} prior knowledge of each other
 {\em nor} an assumption for the WIMP velocity distribution.
\section{Summary and conclusions}
 In this article
 I reviewed the methods for the determination(s) of the mass
 (and eventually the spin--independent cross section on nucleons)
 of Weakly Interacting Massive Particles
 with positive signals of their elastic scattering off target nuclei
 in direct Dark Matter detection experiments.

 With only one experiment,
 the WIMP mass combined with its SI cross section on nucleons
 could be estimated by the maximum likelihood analysis
 using a theoretically predicted scattering spectrum
 fitted to the measured recoil energies. 
 If WIMPs are light ($\mchi < \mN$),
 the shape of the recoil spectrum is sensitive to their mass,
 then the WIMP mass (and also the cross section)
 can be estimated with a higher accuracy;
 however,
 in case WIMPs are (much) heavier than the target nucleus
 ($\mchi \gsim 200$ GeV),
 the recoil spectrum becomes nearly independent on $\mchi$ and
 it is then more difficult or even impossible
 to estimate the WIMP mass reasonably
 with $\cal O$(100) events.

 The maximum likelihood analysis depends on
 the prior assumption for the velocity distribution of halo WIMPs
 as well as on the local WIMP density.
 For a WIMP mass of 100 GeV,
 an $\sim$ 10\% measurement uncertainty on
 the orbital velocity of the Solar system
 could cause an $\sim$ 20\% systematic error on
 the best--fit WIMP mass combined
 with an $\sim$ 10\% error on
 the SI WIMP--proton cross section.
 
 In order to determine the WIMP mass
 without making any assumption for
 the WIMP velocity distribution,
 I described a second method based on
 the reconstruction of (the moments of)
 the WIMP velocity distribution function
 from two experiments with different target nuclei.
 This method can be used without knowing
 the WIMP--nucleus cross section.
 The only information needed is the measured recoil energies.
 By matching the maximal cut--off energies of two experiments
 one could in principle estimate the WIMP mass up to $\sim$ 500 GeV
 with $\cal O$(50) events from each experiment.

 Nevertheless,
 the algorithmic procedure for determining
 the maximal cut--off energy of the experiment
 with the lighter target nucleus by minimizing $\chi^2$
 could overestimate the WIMP mass by 15 to 20\%
 if WIMPs are light,
 or lead to unreliable error estimates
 if WIMPs are heavy.
 The latter could become worse with larger event samples.
 However,
 the fact that
 optimal $\Qmax$ matching works well in all cases,
 for both the median reconstructed WIMP mass and its statistical error,
 gives us hope that
 a better algorithm for $\Qmax$ matching can be found
 which only relies on the data.

 Additionally,
 by combining two (or three) experimental data sets
 one could also estimate the spin--independent WIMP--proton coupling
 without knowing the WIMP mass.
 Although,
 due to the degeneracy between
 the local WIMP density and the WIMP-nucleus cross section,
 one needs to adopt the local Dark Matter density
 (as the unique assumption),
 at least an upper bound on this coupling could be given.
 In fact,
 for a WIMP mass of 100 GeV,
 with $\cal O$(50) events from each experiment,
 a statistical uncertainty of $\sim$ 15\% could be reached.
 This is much smaller than
 the systematic uncertainty on the local Dark Matter density
 (of a factor of 2 or even larger).

 In summary,
 by means of currently running and projected experiments
 using detectors with $10^{-9}$ to $10^{-11}$ pb sensitivities
 \cite{{Baudis07a}, {Drees08}, {Bednyakov08}}
 (see footnote \ref{NJP-DM-Part3}),
 we stand a good chance of detecting Dark Matter particles,
 if Dark Matter indeed consists (mainly) of WIMPs.
 Then the methods presented here
 can be used to estimate the mass
 (and eventually the cross section on nucleons)
 of Dark Matter particles.
 This information
 (perhaps combined with information
  from indirect detection experiments
  \cite{Bernal08})
 will allow us
 not only to constrain the parameter space
 in different extensions of the Standard Model of particle physics,
 but also to identify WIMPs among new particles produced at colliders
 (hopefully in the near future).
 Once one is confident of this identification,
 one can use further collider measurements of the mass and couplings of WIMPs.
 Together with the reconstruction of
 the velocity distribution of halo WIMPs \cite{DMDDf1v},
 this will then yield a new determination of the local WIMP density.
 On the other hand,
 knowledge of the WIMP couplings will also
 permit prediction of the WIMP annihilation cross section.
 Together with information on the WIMP density,
 this will allow one to predict the event rate
 in the indirect Dark Matter detection
 \cite{{SUSYDM96}, {Bertone05}}
 as well as to test
 our understanding of the early Universe.
\subsubsection*{Acknowledgments}
 The author would like to thank M.~Drees and A.~M.~Green
 for detailed comments on the preliminary draft.
 The author also appreciates IOP Publishing Limited
 for their kind permission to reproduce published plots
 in this article.
 This work was partially supported
 by the BK21 Frontier Physics Research Division under project
 no.~BA06A1102 of Korea Research Foundation.
\appendix
\setcounter{equation}{0}
\setcounter{figure}{0}
\renewcommand{\theequation}{A\arabic{equation}}
\renewcommand{\thefigure}{A\arabic{figure}}
%
%
\section{Formulae needed in Sec.~3}
 Here I list all formulae needed
 in the model--independent method described in Sec.~3.
 Detailed derivations and discussions
 can be found in Refs.~\cite{{DMDDf1v}, {DMDDmchi}}.
\subsection{Estimating \boldmath$r(\Qmin)$, $I_n(\Qmin, \Qmax)$,
            and their statistical errors}
 First,
 consider experimental data described by
\beq
     {\T Q_n - \frac{b_n}{2}}
 \le \Qni
 \le {\T Q_n + \frac{b_n}{2}}
\~,
     ~~~~~~~~~~~~ 
     i
 =   1,~2,~\cdots,~N_n,~
     n
 =   1,~2,~\cdots,~B.
\label{eqn:Qni}
\eeq
 Here the total energy range between $\Qmin$ and $\Qmax$
 has been divided into $B$ bins
 with central points $Q_n$ and widths $b_n$.
 In each bin,
 $N_n$ events will be recorded.
 Since the recoil spectrum $dR/dQ$ is expected
 to be approximately exponential,
 the following ansatz for the spectrum in the $n$th bin
 has been introduced \cite{DMDDf1v}:
\beq 
        \adRdQ_n
 \equiv \adRdQ_{Q \simeq Q_n}
 \equiv \rn  \~ e^{k_n (Q - Q_{s, n})}
\~.
\label{eqn:dRdQn}
\eeq
 Here $r_n$ is the standard estimator for $dR/dQ$ at $Q = Q_n$:
\beq
   r_n
 = \frac{N_n}{b_n}
\~,
\label{eqn:rn}
\eeq
 $k_n$ is the logarithmic slope of the recoil spectrum in the $n$th bin,
 which can be computed numerically
 from the average $Q-$value in the $n$th bin:
\beq
   \bQn
 = \afrac{b_n}{2} \coth\afrac{k_n b_n}{2}-\frac{1}{k_n}
\~,
\label{eqn:bQn}
\eeq
 where
\beq
        \bQxn{\lambda}
 \equiv \frac{1}{N_n} \sumiNn \abrac{\Qni - Q_n}^{\lambda}
\~.
\label{eqn:bQn_lambda}
\eeq
 The error on the logarithmic slope $k_n$
 can be computed from Eq.(\ref{eqn:bQn}) directly:
\beq
   \sigma^2(k_n)
 = k_n^4
   \cbrac{  1
          - \bfrac{k_n b_n / 2}{\sinh (k_n b_n / 2)}^2}^{-2}
            \sigma^2\abrac{\bQn}
\~,
\label{eqn:sigma_kn}
\eeq
 with
\beq
   \sigma^2\abrac{\bQn}
 = \frac{1}{N_n-1} \bbigg{\bQQn - \bQn^2}
\~.
\label{eqn:sigma_bQn}
\eeq
 $Q_{s,n}$ in the ansatz (\ref{eqn:dRdQn}) is the shifted point at which
 the leading systematic error due to the ansatz
 is minimal \cite{DMDDf1v},
\beq
   Q_{s, n}
 = Q_n + \frac{1}{k_n} \ln\bfrac{\sinh(k_n b_n/2)}{k_n b_n/2}
\~.
\label{eqn:Qsn}
\eeq
 Note that $Q_{s, n}$ differs from the central point of the $n$th bin, $Q_n$.
 From the ansatz (\ref{eqn:dRdQn}),
 the counting rate at $Q = \Qmin$ can be calculated by
\beq
   r(\Qmin)
 = r_1 e^{k_1 (\Qmin - Q_{s, 1})}
\~,
\label{eqn:rmin_eq}
\eeq
 and its statistical error can be expressed as
\beq
   \sigma^2(r(\Qmin))
 = r^2(\Qmin) 
   \cbrac{  \frac{1}{N_1}
          + \bbrac{  \frac{1}{k_1}
                   - \afrac{b_1}{2} 
                     \abrac{  1
                            + \coth\afrac{b_1 k_1}{2}}}^2
            \sigma^2(k_1)}
\~,
\label{eqn:sigma_rmin}
\eeq
 since
\beq
   \sigma^2(r_n)
 = \frac{N_n}{b_n^2}
\~.
\label{eqn:sigma_rn}
\eeq
 Finally,
 since all $I_n$ are determined from the same data,
 they are correlated with
\beq
   {\rm cov}(I_n, I_m)
 = \sum_a \frac{Q_a^{(n+m-2)/2}}{F^4(Q_a)}
\~,
\label{eqn:cov_In}
\eeq
 where the sum again runs over all events
 with recoil energy between $\Qmin$ and $\Qmax$. 
 And the correlation between the errors on $r(\Qmin)$,
 which is calculated entirely from the events in the first bin,
 and on $I_n$ is given by
\beqn
 \conti {\rm cov}(r(\Qmin), I_n)
        \non\\
 \=     r(\Qmin) \~ I_n(\Qmin, \Qmin + b_1)
        \non\\
 \conti ~~~~ \times 
        \cleft{  \frac{1}{N_1} 
               + \bbrac{  \frac{1}{k_1}
                        - \afrac{b_1}{2} \abrac{1 + \coth\afrac{b_1 k_1}{2}}}}
        \non\\
 \conti ~~~~~~~~~~~~~~ \times 
        \cright{ \bbrac{  \frac{I_{n+2}(\Qmin, \Qmin + b_1)}
                               {I_{n  }(\Qmin, \Qmin + b_1)}
                        - Q_1
                        + \frac{1}{k_1}
                        - \afrac{b_1}{2} \coth\afrac{b_1 k_1} {2}}
        \sigma^2(k_1)}
\~;
\label{eqn:cov_rmin_In}
\eeqn
 note that
 the sums $I_i$ here only count in the first bin,
 which ends at $Q = \Qmin + b_1$.

 On the other hand,
 with a functional form of the recoil spectrum
 (e.g., fitted to experimental data),
 $(dR/dQ)_{\rm expt}$,
 one can use the following integral forms
 to replace the summations given above.
 Firstly,
 the average $Q-$value in the $n$th bin
 defined in Eq.(\ref{eqn:bQn_lambda})
 can be calculated by
\beq
   \bQxn{\lambda}
 = \frac{1}{N_n} \intQnbn \abrac{Q - Q_n}^{\lambda} \adRdQ_{\rm expt} dQ
\~.
\label{eqn:bQn_lambda_int}
\eeq
 For $I_n(\Qmin, \Qmax)$ given in Eq.(\ref{eqn:In_sum}),
 we have
\beq
   I_n(\Qmin, \Qmax)
 = \int_{\Qmin}^{\Qmax} \frac{Q^{(n-1)/2}}{F^2(Q)} \aDd{R}{Q}_{\rm expt} dQ
\~,
\label{eqn:In_int}  
\eeq 
 and similarly for the covariance matrix for $I_n$
 in Eq.(\ref{eqn:cov_In}),
\beq
   {\rm cov}(I_n, I_m)
 = \int_{\Qmin}^{\Qmax} \frac{Q^{(n+m-2)/2}}{F^4(Q)} \aDd{R}{Q}_{\rm expt} dQ
\~.
\label{eqn:cov_In_int}
\eeq 
 Remind that
 $(dR/dQ)_{\rm epxt}$ is the {\em measured} recoil spectrum
 {\em before} the normalization by the exposure.
 Finally,
 $I_i(\Qmin, \Qmin + b_1)$ needed in Eq.(\ref{eqn:cov_rmin_In})
 can be calculated by
\beq
   I_n(\Qmin, \Qmin + b_1)
 = \int_{\Qmin}^{\Qmin + b_1}
   \frac{Q^{(n-1)/2}}{F^2(Q)} \bbigg{r_1 \~ e^{k_1 (Q - Q_{s, 1})}} dQ
\~.
\label{eqn:In_1_int}  
\eeq 
 Note that $r(\Qmin)$ and $I_n(\Qmin, \Qmin + b_1)$ 
 should be estimated by Eqs.(\ref{eqn:rmin_eq}) and (\ref{eqn:In_1_int})
 with $r_1$, $k_1$ and $Q_{s, 1}$
 estimated by Eqs.(\ref{eqn:rn}), (\ref{eqn:bQn}), and (\ref{eqn:Qsn})
 in order to use the other formulae for estimating
 the (correlations between the) statistical errors
 without any modification.
\subsection{Statistical errors on \boldmath$\mchi$
            given in Eqs.(\ref{eqn:mchi_Rn}) and (\ref{eqn:mchi_Rsigma})}
 The expression for $\left. \mchi \right|_{\Expv{v^n}}$
 given in Eq.(\ref{eqn:mchi_Rn})
 leads to a lengthy expression for its statistical error:
\beqn
        \left. \sigma(\mchi) \right|_{\Expv{v^n}}
 \=     \frac{\sqrt{\mX / \mY} \vbrac{\mX - \mY} \abrac{\calR_{n, X} / \calR_{n, Y}} }
             {\abrac{\calR_{n, X} / \calR_{n, Y} - \sqrt{\mX / \mY}}^2}
        \non\\
 \conti ~~~~ \times
        \bbrac{  \frac{1}{\calR_{n, X}^2}
                 \sum_{i, j = 1}^3
                 \aPp{\calR_{n, X}}{c_{i,X}} \aPp{\calR_{n, X}}{c_{j,X}}
                 {\rm cov}(c_{i,X}, c_{j,X})
               + (X \lto Y)}^{1/2}
\!.
\label{eqn:sigma_mchi_Rn}
\eeqn
 Here a short--hand notation for the six quantities
 on which the estimate of $\mchi$ depends has been introduced:
\beq
   c_{1, X}
 = I_{n, X}
\~,
    ~~~~~~~~~~~~ 
   c_{2, X}
 = I_{0, X}
\~,
   ~~~~~~~~~~~~ 
   c_{3, X}
 = r_X(Q_{{\rm min}, X})
\~;
\label{eqn:ciX}
\eeq
 and similarly for the $c_{i, Y}$.
 Estimators for ${\rm cov}(c_i, c_j)$ have been given
 in Eqs.(\ref{eqn:cov_In}) and (\ref{eqn:cov_rmin_In}).
 Explicit expressions for the derivatives of $\calR_{n, X}$
 with respect to $c_{i, X}$ are:
\cheqnXa{A}
\beq
   \Pp{\calR_{n, X}}{\InX}
 = \frac{n+1}{n}
   \bfrac{  F^2_X(Q_{{\rm min}, X})}
         {  2 Q_{{\rm min}, X}^{(n+1)/2} r_X(Q_{{\rm min}, X})
          + (n+1) \InX F^2_X(Q_{{\rm min}, X})}
   \calR_{n, X}
\~,
\label{eqn:dRnX_dInX}
\eeq
\cheqnXb{A}
\beq
   \Pp{\calR_{n, X}}{\IzX}
 =-\frac{1}{n}
   \bfrac{  F^2_X(Q_{{\rm min}, X})}
         {  2 Q_{{\rm min}, X}^{1/2} r_X(Q_{{\rm min}, X})
          + \IzX F^2_X(Q_{{\rm min}, X})}
   \calR_{n, X}
\~,
\label{eqn:dRnX_dIzX}
\eeq
 and
\cheqnXc{A}
\beqn
        \Pp{\calR_{n, X}}{r_X(Q_{{\rm min}, X})}
 \=     \frac{2}{n}
        \bfrac{    Q_{{\rm min}, X}^{(n+1)/2} \IzX - (n+1) Q_{{\rm min}, X}^{1/2} \InX}
              {  2 Q_{{\rm min}, X}^{(n+1)/2} r_X(Q_{{\rm min}, X})
               + (n+1) \InX F^2_X(Q_{{\rm min}, X})}
        \non\\
 \conti ~~~~~~~~~~~~~~~~ \times 
        \bfrac{  F^2_X(Q_{{\rm min}, X})}
              {  2 Q_{{\rm min}, X}^{   1 /2} r_X(Q_{{\rm min}, X})
               +       \IzX F^2_X(Q_{{\rm min}, X})}
        \calR_{n, X}
\~;
\label{eqn:dRnX_drminX}
\eeqn
\cheqnX{A}%
 explicit expressions for the derivatives of $\calR_{n, Y}$
 with respect to $c_{i, Y}$ can be given analogously.
 Note that,
 firstly,
 factors $\calR_{n, (X, Y)}$ appear in all these expressions,
 which can practically be cancelled by the prefactors
 in the bracket in Eq.(\ref{eqn:sigma_mchi_Rn}).
 Secondly,
 all the $\IzX,~\IzY,~\InX,~\InY$ should be understood
 to be computed according to Eqs.(\ref{eqn:In_sum}) or (\ref{eqn:In_int})
 with integration limits $\Qmin$ and $\Qmax$ specific for that target.

 Similar to the analogy between Eqs.(\ref{eqn:mchi_Rn}) and (\ref{eqn:mchi_Rsigma}),
 the statistical error on $\left. \mchi \right|_\sigma$
 given in Eq.(\ref{eqn:mchi_Rsigma})
 can be expressed as
\beqn
        \left. \sigma(\mchi) \right|_\sigma
 \=     \frac{\abrac{\mX / \mY}^{5/2} \vbrac{\mX - \mY}
              \abrac{\calR_{\sigma, X} / \calR_{\sigma, Y}} }
             {\bbrac{\calR_{\sigma, X} / \calR_{\sigma, Y} - \abrac{\mX / \mY}^{5/2}}^2}
        \non\\
 \conti ~~~~~~ \times 
        \bbrac{  \frac{1}{\calR_{\sigma, X}^2}
                 \sum_{i, j = 2}^3
                 \aPp{\calR_{\sigma, X}}{c_{i,X}} \aPp{\calR_{\sigma, X}}{c_{j,X}}
                 {\rm cov}(c_{i,X}, c_{j,X})
               + (X \lto Y)}^{1/2}
\~,
\label{eqn:sigma_mchi_Rsigma}
\eeqn
 where we have again used the short--hand notation in Eq.(\ref{eqn:ciX});
 note that $c_{1, (X, Y)} = I_{n, (X, Y)}$ do not appear here.
 Expressions for the derivatives of $\calR_{\sigma, X}$
 can be computed from Eq.(\ref{eqn:Rsigma_min}) as
\cheqnXa{A}
\beq
   \Pp{\calR_{\sigma, X}}{\IzX}
 = \bfrac{  F_X^2(Q_{{\rm min}, X})}
         {  2 Q_{{\rm min}, X}^{1/2} r_X(Q_{{\rm min}, X})
          + \IzX F_X^2(Q_{{\rm min}, X})}
   \calR_{\sigma, X}
\~,
\label{eqn:dRsigmaX_dIzX}
\eeq
\cheqnXb{A}
\beq
   \Pp{\calR_{\sigma, X}}{r_X(Q_{{\rm min}, X})}
 = \bfrac{  2 Q_{{\rm min}, X}^{1/2}} 
         {  2 Q_{{\rm min}, X}^{1/2} r_X(Q_{{\rm min}, X})
          + \IzX F_X^2(Q_{{\rm min}, X})}
   \calR_{\sigma, X}
\~;
\label{eqn:dRsigmaX_drminX}
\eeq
\cheqnX{A}
 and similarly for the derivatives of $\calR_{\sigma, Y}$.
\subsection{Covariance of \boldmath$f_i$
            defined in Eqs.(\ref{eqn:fiXa}) and (\ref{eqn:fiXb})}
 The entries of the $\cal C$ matrix in Eq.(\ref{eqn:Cij})
 involving basically only the moments of the WIMP velocity distribution
 can be read off Eq.(82) of Ref.~\cite{DMDDf1v},
 with an slight modification
 due to the normalization factor in Eq.(\ref{eqn:fiXa})%
\footnote{
 Since the last $f_i$ defined in Eq.(\ref{eqn:fiXb})
 can be computed from the same basic quantities,
 i.e., the counting rates at $\Qmin$ and the integrals $I_0$,
 it can directly be included in the covariance matrix.
}:
\beqn
        {\rm cov}\abrac{f_i, f_j}
 \=     \calN_{\rm m}^2
        \bbiggl{  f_i \~ f_j \~ {\rm cov}(I_0, I_0)
                + \Td{\alpha}^{i+j} (i+1) (j+1) {\rm cov}(I_i, I_j)}
        \non\\
 \conti ~~~~~~~~ 
                - \Td{\alpha}^j (j+1) f_i \~ {\rm cov}(I_0, I_j)
                - \Td{\alpha}^i (i+1) f_j \~ {\rm cov}(I_0, I_i)\bigg.
        \non\\
 \conti ~~~~~~~~~~~~ 
                + D_i D_j \sigma^2(r(\Qmin))
                - \abrac{D_i f_j + D_j f_i} {\rm cov}(r(\Qmin), I_0)\Bigg.
        \non\\
 \conti ~~~~~~~~~~~~~~~~ 
        \bbiggr{+ \Td{\alpha}^j (j+1) D_i \~ {\rm cov}(r(\Qmin), I_j)
                + \Td{\alpha}^i (i+1) D_j \~ {\rm cov}(r(\Qmin), I_i)}
\~.
        \non\\
\label{eqn:cov_fi}
\eeqn
 Here
\beq
        \calN_{\rm m}
 \equiv \frac{1}{2 \Qmin^{1 /2} r(\Qmin)/\FQmin + I_0}
\~,
\label{eqn:calNm}
\eeq
\beq
        \Td{\alpha}
 \equiv \frac{\alpha}{300~{\rm km/s}}
\~,
\label{eqn:td_alpha}
\eeq
 and
\cheqnXa{A}
\beq
        D_i
 \equiv \frac{1}{\cal N_{\rm m}} \bPp{f_i}{r(\Qmin)}
 =      \frac{2}{\FQmin}
        \abigg{\Td{\alpha}^i \Qmin^{(i+1)/2} - \Qmin^{1/2} \~ f_i}
\~,
\label{eqn:Dia}
\eeq
 for $i = -1,~1,~2,~\dots,~n_{\rm max}$; and
\cheqnXb{A}
\beq
   D_{n_{\rm max}+1}
 = \frac{2}{\FQmin} \abrac{-\Qmin^{1/2} f_{n_{\rm max}+1}}
\~.
\label{eqn:Dib}
\eeq
\cheqnX{A}
\subsection{Statistical error on \boldmath$|f_{\rm p}|^2$
            given in Eq.(\ref{eqn:rho0_fp2})}
 From Eq.(\ref{eqn:rho0_fp2}),
 it can easily be found that
\beq
    \sigma(|f_{\rm p}|^2)
 = |f_{\rm p}|^2
    \bbrac{  \frac{\sigma^2(\mchi)}{(\mchi + \mN)^2}
           + \calN_{\rm m}^2 \~ \sigma^2(1 / \calN_{\rm m})
           + \frac{2 \calN_{\rm m} \~ {\rm cov}(\mchi, 1 / \calN_{\rm m})}{(\mchi + \mN)}}^{1/2}
\~,
\label{eqn:sigma_fp2}
\eeq
 where $\calN_{\rm m}$ is defined in Eq.(\ref{eqn:calNm}),
 and
\beq
      \sigma^2(1 / \calN_{\rm m})
 =    \bfrac{2 \Qmin^{1/2}}{\FQmin}^2 \sigma^2(r(\Qmin))
  +   \sigma^2(I_0)
  + 2 \bfrac{2 \Qmin^{1/2}}{\FQmin} {\rm cov}(r(\Qmin), I_0)
\~.
\label{eqn:sigma_calNm}
\eeq
 The correlation between the error on the reconstructed $\mchi$
 and that on the estimator of $1 / \calN_{\rm m}$,
 the third term in Eq.(\ref{eqn:sigma_fp2}),
 can be neglected in case
 one uses three independent data sets.
\end{document}